\documentclass[a4paper,12pt]{article}
\usepackage[utf8]{inputenc}
\usepackage[english]{babel}
\usepackage{authblk}
\usepackage{graphicx}
\usepackage[singlespacing]{setspace}
\usepackage[headheight=1in,margin=1in]{geometry}
\usepackage{fancyhdr}
\usepackage{lipsum}
\usepackage{hyperref}
\usepackage{xcolor}
\usepackage{amsfonts}
\usepackage{amsmath}
\usepackage{latexsym}
\usepackage{caption}
\usepackage[labelfont=bf]{caption}
\usepackage{makecell}
\usepackage{url}

\usepackage{natbib}
\setcitestyle{authoryear,open={(},close={)}} 

\date{}

\begin{document}
\title{Toward a Hybrid Digital Twin of Society: \\ Quantifying Cognitive–Spatial Linkages Through Online–Offline Feedback Networks}
\author{
Rafiazka Hilman,\textsuperscript{1,2}, Júlia Koltai\textsuperscript{1,3}
}

\maketitle

\begin{center}
{
1. MTA–TK Lendület “Momentum” Digital Social Science Research Group for Social Stratification, ELTE Centre for Social Sciences, Budapest, Hungary\\
2. Informatics Institute, University of Amsterdam, Amsterdam, The Netherlands\\
3. Institute of Empirical Studies, ELTE Faculty of Social Sciences, Budapest, Hungary
}\\
\end{center}

\vspace{10mm}

\begin{abstract}
\fontsize{9pt}{11pt}\selectfont 
Digital platforms increasingly shape how people experience and navigate cities, creating a link between virtual information seeking and physical mobility. Despite this growing interdependence, online and offline activities are often studied separately in urban mobility research. This paper proposes the concept and computational framework of the \textit{Feedback Network} to capture the continuous interaction between cognitive activity in digital environments and  behavior in physical space. Using Google Search and Location History data of the same people collected through a data donation framework in Budapest, Hungary about the period, between 2018 and 2022, it examines how online search patterns and offline visitation behavior co-evolve over time. To investigate these dynamics, semantic and spatial analytical approaches are combined. Radius of gyration is adapted to measure the variation in both geographic mobility and semantic exploration, enabling comparison between physical movement and online cognitive dispersion. In parallel, a Feedback Network is developed to model transitions between search related and location related activity clusters. The framework is evaluated with \textit{Concentration Entropy}, which measures the extent to which interconnected behavioral flows are concentrated around routine pathways or distributed across exploratory transitions.

The results indicate that online exploration is systematically more concentrated than offline mobility, suggesting that digital behavior is shaped by narrower and more repetitive semantic interests, whereas physical movement remains relatively more diverse. Persistent linkages are observed between search and visitation activities associated with retail as well as business and professional services, highlighting the presence of stable cognitive-spatial behavioral loops. The analysis also shows that the COVID-19 pandemic disrupted spatial routines more strongly than cognitive exploration, temporarily widening the gap between digital engagement and realized movement. Overall, the findings demonstrate that urban mobility relies on the interaction between informational exposure and spatial encounter rather than through physical movement alone. The proposed Feedback Network offers a foundation for future behavioral Digital Twin frameworks capable of integrating mobility and digital behavior into a unified representation of urban dynamics, conceptualized as \textit{a Hybrid Digital Twin of Society}.
\end{abstract}

\clearpage

\section{Introduction}
\label{Introduction}
Contemporary urban behavior and social dynamics overlap simultaneously across two interconnected yet analytically distinct domains: an offline visit space, defined by the geographic extent of individuals’ physical activities and destination encounters, and an online search space, defined by the semantic breadth of digital information seeking behavior expressed through search queries, keywords, and virtual exploration. Despite growing recognition that digital platforms increasingly mediate how individuals perceive and navigate cities, these two behavioral spaces are still commonly analyzed separately.

The analytical distance in between the two spaces obscures an emerging form of cognitive–spatial dualism in contemporary urban systems. Online search activity frequently precedes offline mobility by revealing latent intentions, preferences, and anticipatory decision making processes, whereas physical encounters often generate subsequent cycles of digital interaction through reviews, navigation requests, and additional information seeking. In the context of daily errands, prospective trips are often preceded by online search activities, such as queries related to restaurants or transport accessibility, which reflect underlying travel intentions and destination preferences. Subsequent physical visits may, in turn, stimulate further digital engagement through activities including expressing experiences in online reviews and finding the shortest route home, thereby illustrating the recursive relationship between digital behavior and human mobility patterns. Consequently, understanding this cognitive–spatial relationship is increasingly important for urban mobility research because cities are no longer experienced solely through physical movement, but also through continuous digital mediation. 

As online platforms have become embedded within everyday mobility practices, the boundary between virtual exposure and physical encounter progressively blurred. Consequently, quantifying the relationship between these two domains provides a foundation for understanding how digital attention shapes urban flows, how informational ecosystems influence mobility behavior, and how cognitive exploration translates into spatial interaction within increasingly hybrid urban environments. As online platforms have become embedded within everyday mobility practices, the boundary between virtual exposure and physical encounter has progressively blurred. Consequently, quantifying the relationship between these two domains provides a foundation for understanding how digital attention shapes urban flows, how informational ecosystems influence mobility behavior, and how cognitive exploration translates into spatial interaction within increasingly hybrid urban environments. In the same manner, examining the reverse relationship provides insight into how physical encounters stimulate subsequent digital engagement, how mobility experiences reshape online information-seeking behavior, and how spatial interaction generates recurring cycles of virtual activity within interconnected urban systems.

This paper aims to address these dynamics by revealing the underlying mechanisms of how digital activity reshapes patterns of co-presence and how physical environment affects digital behavior within cities. To explore this concept, we pose two central research questions. First, are there any similarities in the variance and dynamic of offline exploration and online search behavior of people when exploring offline visit spaces versus online search spaces, and how does this shape their overall mobility experience? Second, how can a continuous behavioral loop be modeled through a cross domain feedback network that links online search behavior (cognitive) with offline visit behavior (spatial)? Our goal is to settle the conceptual framework of the \textit{Feedback Network}, measured with computational approaches that highlight behavioral push dynamics. With regard to this aspect, the current concept paper lays the foundations of this topic of growing importance.

\subsection{Digital Exposure, Physical Encounter, and Behavioral Reinforcement}
\label{Digital Exposure, Physical Encounter, and Behavioral Reinforcement}
Urban mobility is increasingly shaped by two interconnected yet analytically distinct layers of behavior comprising digital exposure, where individuals encounter information through online search and virtual exploration, and physical encounter, where those intentions materialize into movement across urban space through visits to destinations. While traditional mobility research has primarily focused on trajectories, flows, and activity locations, contemporary urban behavior is now deeply mediated by digital infrastructures that influence how people perceive, evaluate, and navigate cities before physical movement even occurs. In this sense, mobility can no longer be understood solely as spatial displacement but rather as a cognitive process embedded within broader informational ecosystems. In this sense, mobility can no longer be understood solely as spatial displacement but rather as a cognitive process embedded within broader informational ecosystems. Conversely, physical location can also shape online search behavior, as being present in a specific place may prompt location-specific queries about nearby services, routes, amenities, or points of interest. This bidirectional relationship highlights how mobility and information seeking jointly structure experience.

Conventional mobility models generally conceptualize movement as an observable outcome driven by spatial constraints such as distance \citep{zhou2026exponential}, accessibility \citep{elorduy2025assessing}, transportation networks \citep{ceder2021urban}, or land use structure \citep{wang2022land}. Foundational studies in human mobility have demonstrated that movement patterns exhibit strong regularities and recurrent structures despite apparent complexity, enabling robust quantitative modeling of travel behavior and activity spaces \citep{gonzalez2008understanding, song2010limits, barbosa2018human}. A number of  works using mobile phone data, GPS trajectories, and large scale geospatial sensing further confirms that urban mobility exhibits stable temporal rhythms, hierarchical visitation structures, and predictable exploration behaviors across cities and populations \citep{pappalardo2015returners, alessandretti2020scales}. These approaches have significantly advanced the understanding of commuting systems, accessibility inequalities, and spatial interaction dynamics. However, capturing realized mobility only does not provides insight into the informational and cognitive processes that precede movement decisions.

While urban mobility research has consistently shown that offline movement follows regular but heterogeneous exploration patterns shaped by daily routines, accessibility, and spatial opportunity structures \citep{gonzalez2008understanding,song2010limits,barbosa2018human}, studies of online attention and information diffusion similarly demonstrate that digital behavior exhibits highly structured yet uneven distributions across social and informational networks \citep{kossinets2008structure,lazer2009social}. In both physical and digital environments, human behavior is neither random nor evenly distributed. Instead, individuals repeatedly concentrate their activity within a limited set of familiar locations \citep{gonzalez2008understanding} or informational environments \citep{gong2024google}, while only occasionally exploring new places or content. This produces strong regularities at the aggregate level alongside substantial variation between individuals and groups.

Search queries, browsing behavior, recommendation systems, and social media engagement often represent early stage cognitive signals associated with destination choice, consumption preference, or activity planning. The rapid integration of mobile platforms into everyday urban life has intensified this relationship, as navigation applications, online reviews, ride hailing systems, and localized search engines continuously mediate how individuals interact with urban space \citep{kitchin2014real, shelton2015actually, qi2024understanding}. Before visiting restaurants, retail centers, transport hubs, or tourist attractions, users commonly engage in online information seeking that shapes where, when, and how mobility is realized. Digital exposure significantly influences spatial visitation patterns, retail attractiveness, and urban service demand, especially in platform mediated urban economies \citep{verhoef2015multi, huang2020twitter, he2026digital}. These findings hint that online behavior should not be interpreted merely as a reflection of offline activity, but rather as an active component of mobility generation itself.

Importantly, the relationship between online and offline behavior is recursive rather than linear. In addition to the above described processes, when digital information search precedes physical movement, physical encounters can also generate new cycles of digital engagement through reviews, social sharing, navigation queries, recommendations, and follow up information seeking. This feedback dynamic creates a continuous loop in which digital and spatial activities mutually reinforce one another. This loop reflects the structural transformation in the nature of urban interaction under conditions of pervasive digital mediation where cities are increasingly experienced simultaneously as physical environments and informational landscapes.

\subsection{Feedback Network as A Framework for Digital Twins}
\label{Feedback Network as A Framework for Digital Twins}
Existing Digital Twin (DT) frameworks have been widely applied to transportation monitoring, infrastructure management, environmental simulation, and mobility optimization because these concepts enable the mirroring of physical processes through continuously updated digital representations \citep{batty2018digital, dembski2020urban, ferre2022adoption}. However, the way DT frameworks are implemented remains primarily infrastructure centric. These applications often focus on observable spatial phenomena such as traffic flows, buildings, road networks, or environmental conditions, while giving comparatively limited attention to the cognitive and behavioral processes that precede and shape urban movement. 

Integrating the interconnection between visit trajectories and the digital traces of online search queries offers an opportunity to extend DT frameworks beyond physical infrastructure toward the behavioral and cognitive dimensions of urban systems such  as Social (Urban) Digital Twin \citep{yossef2023social} or Digital Twin of Society \citep{chircu2023towards}. In contrast, computational social science and urban informatics research increasingly demonstrate that digital traces such as web searches, online queries, and other semantic activity patterns provide meaningful signals of collective behavior and urban demand \citep{lazer2009social, vaccari2009holistic, zinman2020utilizing}. Yet, digital cognition and physical mobility, are still largely analyzed independently, which eventually overlooks the detection of overlaps in spatial cognition and opportunity structures \citep{song2010limits, louail2014mobile, brelsford2019using}. 

The missing component between these domains is a Feedback Network, defined as the continuous behavioral loop connecting cognitive activity in digital space with movement activity in physical space. Given this, a more comprehensive representation of how individuals interact with urban environments across both virtual and physical domains. For instance, the emergence of park popularity is associated with intensified online attention driven by timely digital recommendations or social media exposure \citep{dong2023spatiotemporal, wei2024deciphering}. Similarly, visits to places in the age of the platform economy \citep{du2022want} stimulate people to engage in additional online searches (platform-mediated place search) and generate user-created content (user-generated destination image). Not only that it highlights experience and expectation, but also reflecting how physical urban configuration is entangled with subsequent informational engagement, a phenomenon denoted as platform urbanism \citep{caprotti2022beyond}.

In this context, DT can be reconceptualized not merely as mirrors of physical infrastructure, but as adaptive sociospatial systems capable of capturing the continuous feedback loop between digital cognition and urban mobility behavior. This step envisions the emergence of \textit{a Hybrid Digital Twin of Society}. It enables the transition from static urban replicas toward adaptive behavioral twins capable of modeling not only where people move, but also why they move and what kinds of information drive that movement. The proposed framework conceptualizes the city as a coupled dual layer system composed of an Online Twin and an Offline Twin. The Online Twin represents collective human intent (if the link goes from the search query to location) and consequences (if the link goes from the location to the search query) through search queries, keyword clusters, and semantic categories derived from Google Trends and search history data. These query clusters serve as digital representations of informational demand, interests, and cognitive attention distributed across the urban population. In parallel, the Offline Twin represents the physical layer of urban interaction through visits to points of interest (POIs), like businesses, transportation hubs, and public spaces. 

The Feedback Network acts as the linkage mechanism between these two twins. Rather than treating online and offline activities as isolated systems, the framework models them as mutually informative processes connected through time constrained linkages. A search query may trigger a physical visit, while a physical visit may subsequently generate additional searches, navigation requests, or online interactions. Mathematically, these interactions are represented as linkage matrices connecting semantic query clusters with spatial destination clusters. These matrices include the edges of the DT, enabling the identification of coupled online–offline flows and the prediction of urban demand linkages over time. Recent advances in semantic embedding models, large scale mobility analytics, and urban computing make such integration increasingly feasible. Transformer based language models now enable the extraction of semantically meaningful representations of collective intent from search activity \citep{wang2022text}, while geospatial clustering and mobility modeling techniques allow robust characterization of urban movement patterns and activity spaces.

Identifying and quantifying linkages between online and offline spaces therefore provides a critical lens for examining how digital activity reshapes patterns of co-presence, accessibility, and cognitive-spatial linkage within cities and vice versa. The contribution of this research lies in integrating online and offline activity spaces within a unified entropic framework namely the \textit{Feedback Network}. By systematically connecting digital and physical behaviors, we bridge the digital and urban domains, advancing both the theoretical understanding of digital–urban interactions and the methodological toolkit for analyzing dualism in urban dynamics in the context of pervasive digitalization.\\

\section{Data and Methods}
\label{Data and Methods}
This paper treats the joint presence of the two spaces (online and offline) as a probability distribution over discrete states (locations or search keywords) with which behavioral complexity can be quantified. Capturing visit and search space interactions enables us to move beyond activity counts to estimate the effective number of routinely occupied choices, for instance those locations and searches that frequently co-occur, which is crucial to compare these domains directly. A location and a search query are connected if they appear within a given time window (e.g.: one hour). In the case of retail customer activity, the one-hour time window is justified, as 55\% of consumers completing a mobile-based search take physical or transactional action within a single hour due to immediate on-the-go intent \citep{thinkwithgoogle, molitor2023digitizing}. These ties and nodes create the \textit{Feedback Networks} at the individual level, which are then aggregated across the sample population. Furthermore, by clustering both the spatial locations and the search queries, \textit{Forward Linkage Probability} matrices are constructed to estimate the probability of visiting a specific spatial cluster (defined by locations) given prior search activity on a query cluster (defined by keywords), and vice versa. This approach let us explore robust interaction types between offline locations and digital search behavior. To refine the interpretation of these matrices, the \textit{Concentration Entropy} index captures the online and offline co-occurrence distribution patterns for each location and search cluster, revealing whether links are widely distributed across numerous clusters or concentrated among just a few.

\subsection{Data}
\label{Data}
The trajectory of offline visits at each individual is constructed based on Google location history data and sequence of search keywords are extracted from Google search history of the same people. Both information are collected under a Donation based Data Collection \citep{breuer2023user} in 2023 in Hungary through the data download packages of participants on a representative sample of Hungarian internet users, supplemented with the necessary ethical approval \citep{kmetty_donacio_2024, koltai2025classifying}. From this sample, we focus on the 90 residents of the capital (Budapest) in the analysis and consider data from 2018 to 2022 since data before 2018 is quite scarce. The distribution of the data for the whole period is provided in Supplementary Materials C. To establish urban spatial context, we consider only visits to places within city boundaries, as well as search activities taking place in the corresponding area.

\subsection{Online Search Space}
\label{Online Search Space}
Google search records collected among Budapest residents in the dataset contain approximately 10 million queries, of which 4.5 million are unique keywords. Considering the large volume of queries, it is difficult to observe topic patterns directly at the collective network level, although such patterns remain visible at the individual network level. Therefore, topic categorization through a clustering approach is preferred to refine the analysis of topic distribution and the underlying behavioral context. To improve semantic consistency and reduce noise, a multistage text cleaning pipeline was implemented. First, all search strings were converted to lowercase and stripped of unnecessary whitespace. URLs embedded in queries were parsed using domain extraction procedures so that web addresses were standardized into domain level representations. Emoji characters, punctuation marks, and non-alphanumeric symbols were removed while preserving accented Hungarian characters. Queries consisting solely of numbers or symbols were discarded. In addition, tokens shorter than two characters were excluded to reduce sparsity and typographical noise. 

This preprocessing stage follows established practices in natural language processing (NLP) for short text normalization and semantic retrieval tasks \citep{manning2008introduction}. Removing duplicate and malformed queries is notably important in large scale behavioral datasets because repeated low information queries can distort density based clustering procedures and increase computational burden. Accordingly, after cleaning, duplicate search strings were removed. Each unique query was assigned a unique identifier that enabled efficient broadcasting of semantic labels back to the full dataset after clustering. This deduplication strategy substantially reduced computational complexity and memory requirements while preserving the original frequency structure of the complete dataset.

\subsubsection{Semantic Embeddings and Dimensionality Reduction}
\label{Semantic Embeddings and Dimensionality Reduction}
After preprocessing the text, we applied a transformer based sentence embedding. Semantic representations of search queries were generated using the multilingual E5 transformer model \citep{wang2024multilingual} implemented through the sentence transformer framework. Transformer based sentence embeddings have become a standard approach for semantic similarity analysis because they capture contextual linguistic relationships more effectively than traditional bag-of-words or TF-IDF approaches \citep{reimers2019sentence, wang2022text}. The E5 model was selected because of its strong multilingual retrieval performance and suitability for short search query representations across multiple languages, including Hungarian. The final embedding matrix represents semantically meaningful vector spaces in which similar queries are located closer together according to cosine similarity.

Because high dimensional transformer embeddings are computationally expensive for clustering and often contain noisy local structures, Uniform Manifold Approximation and Projection (UMAP) \citep{healy2024uniform} was applied to reduce embedding dimensionality prior to clustering. The UMAP model was configured with 30 nearest neighbors and a 10-dimensional latent representation using cosine distance metrics. This optimal configuration resulted from a parameter sweep that considered the trade-off between obtaining a reasonable number of clusters and preserving the semantic similarity of queries based on embedding distance. UMAP is widely used for large scale text embedding reduction because it preserves both local and global manifold structures while maintaining computational scalability \citep{mcinnes2018umap}. Compared with alternatives such as t-SNE, UMAP provides better preservation of global semantic organization and substantially faster runtime performance for millions of observations.

\subsubsection{Cluster Formation}
\label{Cluster Formation}
Clusters of semantically similar search queries were identified using Hierarchical Density based Spatial Clustering of Applications with Noise (HDBSCAN) \citep{mcinnes2017hdbscan}. Following an iterative hyperparameter tuning process across various combinations, the final clustering model was implemented using Euclidean distance in the reduced UMAP space, with a minimum cluster size of 80 observations and a minimum sample threshold of 20. This configuration was chosen to yield an optimal number of clusters, striking a reasonable tradeoff between cluster size and semantic similarity. HDBSCAN was selected because it can identify clusters with varying density distributions while simultaneously detecting outliers and noise points \citep{campello2013density}. This characteristic is advantageous for digital search behavior datasets, where search activity exhibits highly uneven semantic distributions and long-tail query structures. Unlike centroid based methods such as k-means, HDBSCAN does not require a predefined number of clusters and is more robust to irregular semantic cluster geometries. Importantly, clustering was performed exclusively on unique queries rather than the full dataset to avoid density distortions caused by repetitive search terms. This approach improved both computational and cluster stability. This step results in 30 stable clusters and 1 noise cluster. The noise cluster was retained as is and later mapped into a meaningful category as well.

To construct reasonable number but still interpretable thematic categories, clusters were mapped onto a predefined taxonomy derived from Google Trends topical classifications that include 25 high level  categories through semantic similarity matching \citep{googleGoogleTrends}. First, category labels were embedded using the same multilingual E5 transformer model applied to user queries. After that, mean embedding vectors were computed for each detected cluster. Cosine similarity scores between cluster centroids and category embeddings were then calculated, and each cluster was assigned to the category with the highest semantic similarity score. At this point, the noise cluster exhibited the closest distance to the most generic category, namely Reference (Cluster 21), hence, it was retained under that category. To validate the assignment, the clusters and their corresponding keywords were evaluated using the GPT-5.5 model to verify labeling accuracy, with the final results demonstrating  agreement.

\begin{figure}[!ht]
  \centering
  \includegraphics[width=1.1\textwidth]{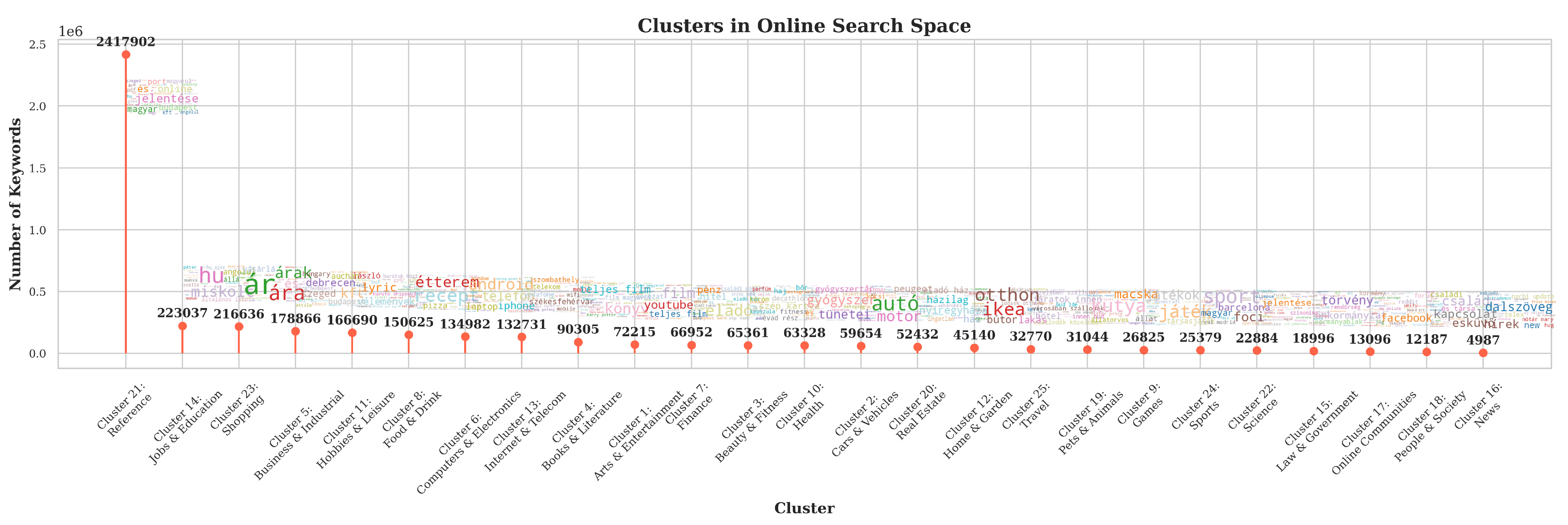}
  \caption{Distribution of online search clusters, derived from the corresponding 25 Google Trends categories.}
  \label{online_cluster}
\end{figure}

This pipeline was chosen for its scalable and language-agnostic classification capabilities when handling unstructured digital trace data. The clustering results are presented in Figure \ref{online_cluster}. Cluster 21 is the largest online cluster by number of keywords, encompassing queries related to general knowledge. As such, it includes searches for word definitions, acronyms, language learning queries, how to guides, quotes, geographical boundaries, and factual data. Rounding out the top five search query topics are Job \& Education (Cluster 14), Shopping (Cluster 23), Business \& Industrial (Cluster 5), and Hobbies \& Leisure (Cluster 11).

\subsection{Offline Visit Space}
\label{Offline Visit Space}
To initialize clusters of location types, for each participant, Google location history and associated place metadata was extracted from Google Timeline activity logs. It consists of 300 thousand individual level place visit records across nearly 18 thousand unique locations in Budapest. The analytical workflow integrates the detection of visit locations, and semantic place classifications to examine urban mobility behavior and activity participation patterns.

As a preprocessing, records with missing or invalid place identifiers (30\%) were excluded to ensure spatial consistency and semantic interpretability. In the dataset, each visit is expected to be tagged with either a location name or geolocation information. However, these attributes are not always available. In such cases, records lacking either form of information are excluded from the dataset. Visit duration was calculated as the temporal difference between arrival and departure timestamps. To remove transient pass through visits and GPS noise, only visits with a minimum duration of five minutes were considered valid activity episodes. Such temporal filtering procedures are commonly used in mobility studies to distinguish meaningful activity participation from short duration movement artifacts \citep{stopher2007assessing}.

To characterize the functional structure of visited places, Google Place identifiers, such as location names (e.g.: Tesco), were first processed via reverse geocoding using OpenStreetMap (OSM) \citep{openstreetmapOpenStreetMap} to retrieve location types. These location types were then mapped to 10 core top-level parent of Foursquare taxonomy labels \citep{foursquareFoursquareCategories}, which were grouped into 10 distinct clusters using the GPT-5.5 model in order to conduct a thematic clustering of the locations that relies heavily on nuanced semantic labels. The grouping based on Foursquare is motivated by the intention to generate a well-defined benchmark for location categorization \citep{spyratos2017using}, while simultaneously enhancing contextual interpretation.

\begin{figure}[!ht]
  \centering
  \includegraphics[width=0.8\textwidth]{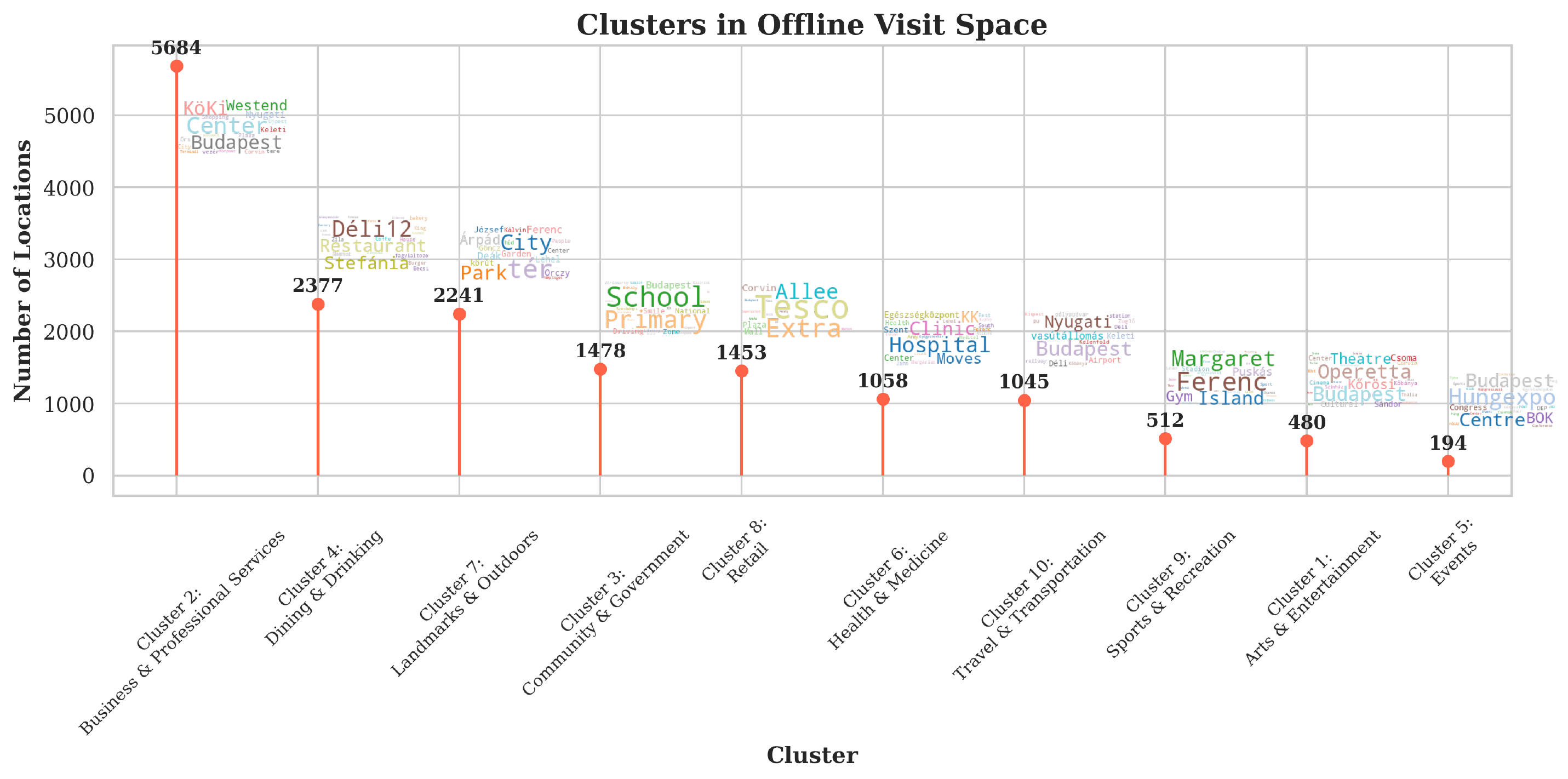}
  \caption{Distribution of offline visit clusters, derived from the corresponding 10 Foursquare categories.}  
  \label{offline_cluster}
\end{figure}

Figure 2 shows the cluster formation within the offline visit spaces. Based on location count, Business \& Professional Services (Cluster 2) is the largest of the ten clusters, accounting for one third of all listed venues. This cluster frequently features common Hungarian corporate legal identifiers, such as 'kft' (\textit{korlátolt felelősségű társaság} / limited liability company), 'bt' (\textit{betéti társaság} / limited partnership), and 'ec' (\textit{egyéni cég} / sole proprietorship). These locations are typically concentrated in central business districts or adjacent to commercial and transit hubs. The next largest categories are Dining \& Drinking (Cluster 4), characterized by identifiers such as 'deli' and 'restaurant', followed closely by Landmarks \& Outdoors (Cluster 7), which commonly include the terms tér (square) and park. The spatial distribution of these locations is provided in Supplementary Materials D.

\subsection{Radius of Gyration}
\label{Radius of Gyration}
To quantify concentration and exploration patterns across both urban mobility and online information behavior, we extend the concept of radius of gyration ($R_g$), a widely used metric in human mobility research that measures the characteristic spatial dispersion of an individual’s trajectory around its center of mass \citep{gonzalez2008understanding, barbosa2018human}. Recalling that we focus parallelism between online and offline activities, the size of exploration in both spaces can be applied as a descriptive measure. Accordingly, the metric is reformulated into two parallel derivatives that reflect the dual nature of human activity namely extension of movement through physical urban space (spatial) and scale of exploration through semantic information space (cognitive). This dual formulation allows us to evaluate whether broader spatial mobility is associated with broader topical exploration in online behavior.

The first derivative corresponds to the conventional physical radius of gyration, which measures the characteristic travel radius of an individual across geographic locations in the city. Let $x_i$ denote the spatial coordinates of visited locations and $w_i$ is the optional visit frequency or dwell time weight associated with each location. The centroid of the trajectory is defined as
\begin{equation}
\mu_x =
\frac{
\sum_{i=1}^{N} w_i x_i
}{
\sum_{i=1}^{N} w_i
}
\end{equation}

and the physical radius of gyration is given by
\begin{equation}
R_g^{spa} =
\sqrt{
\frac{
\sum_{i=1}^{N}
w_i
\left\| x_i - \mu_x \right\|^2
}{
\sum_{i=1}^{N} w_i
}
}
\end{equation}

where $\left\| x_i - \mu_x \right\|$ represents the geographic distance between each visited location and the individual’s spatial centroid. A larger $R_g^{phys}$ indicates that an individual travels across a wider urban area, whereas a smaller value reflects concentrated daily mobility centered around a limited activity space.

The second derivative introduces a semantic radius of gyration, designed to capture topical dispersion in online search behavior. Instead of geographic coordinates, each search query is represented as a semantic embedding vector $v_i \in \mathbb{R}^d$, generated using multilingual E5 transformer embeddings. In this semantic space, distances between vectors reflect topical cosine similarity rather than physical proximity. Queries discussing similar themes occupy nearby positions, while semantically unrelated topics are located farther apart.

The semantic centroid is defined as
\begin{equation}
\mu_v =
\frac{
\sum_{i=1}^{N} w_i v_i
}{
\sum_{i=1}^{N} w_i
}
\end{equation}

The semantic radius of gyration is computed as
\begin{equation}
R_g^{cog} =
\sqrt{
\frac{
\sum_{i=1}^{N}
w_i
\left\| v_i - \mu_v \right\|^2
}{
\sum_{i=1}^{N} w_i
}
}
\end{equation}

where $\left\| v_i - \mu_v \right\|$ measures semantic distance within the embedding space. In this interpretation, a small semantic radius indicates highly concentrated interests centered around a limited set of related topics, while a large semantic radius reflects broad thematic exploration across heterogeneous domains.

For example, a user repeatedly searching for 'Budapest transit pass' and 'metro station' would generate semantically similar embeddings clustered within transportation related regions of the vector space, resulting in a relatively small $R_g^{sem}$. In contrast, a user searching for 'stock market', 'goulash recipe', `Puskas football stadium', and 'tax regulations' would produce embeddings distributed across distant semantic regions, leading to a substantially larger semantic radius of gyration.

Conceptually, the two metrics capture parallel forms of exploration. Physical radius of gyration reflects the spatial breadth of urban mobility, whereas semantic radius of gyration measures the cognitive or informational breadth of online activity. By jointly examining these two dimensions, the framework enables the investigation of whether geographically exploratory individuals also demonstrate broader thematic exploration online, thereby linking mobility behavior with patterns of digital attention and urban information consumption.

\subsection{Cognitive-Spatial Feedback Network}
\label{Cognitive-Spatial Feedback Network}
The \textit{Feedback Network} is modeled as a directed weighted graph that captures temporal linkages between online search activities and offline visit activities. For each individual \(u\), let

\begin{equation}
\mathcal{E}_u =
\left\{
e_1, e_2, \dots, e_n
\right\}
\end{equation}

denote the chronologically ordered sequence of activities, where each event is represented as
\begin{equation}
e_i = (t_i, a_i, c_i).
\end{equation}

Here, \(t_i\) denotes the timestamp of event \(i\), \(a_i \in \{\text{online}, \text{offline}\}\) indicates whether the activity belongs to the online search domain or offline visit domain, and \(c_i \in \mathcal{C}\) represents the above described semantic or spatial cluster associated with the activity. These events create the nodes of the network.

To operationalize the cognitive-spatial feedback process, only heterogeneous linkages are retained. In other words, edges are constructed exclusively between activities belonging to different behavioral domains, so between online queries and offline visits:

\begin{equation}
\text{online} \rightarrow \text{offline},
\qquad
\text{offline} \rightarrow \text{online}.
\end{equation}

A directed edge, starting from the event (search query or visit place) with earlier timestamp and heading towards the event with the later timestamp, is created when the events occur within a predefined temporal threshold \(\Delta t\):

\begin{equation}
0 < (t_j - t_i) \leq \Delta t,
\qquad
a_i \neq a_j.
\end{equation}

In case of events, not the original search queries or visit places, but the clusters where they belong to are considered. Therefore, the Feedback Network for individual \(u\) is therefore defined as

\begin{equation}
G_u = (V_u, W_u),
\end{equation}

where \(V_u\) is the set of semantic and spatial clusters observed in the activity sequence, and \(W_u\) is the weighted adjacency matrix describing linkage magnitude between clusters. The edge weight between cluster \(c_p\) and cluster \(c_q\) is computed as the number of valid heterogeneous linkages:

\begin{equation}
w_{pq}
=
\sum_{i<j}
\mathbf{1}
\left[
c_i = c_p,
\;
c_j = c_q,
\;
a_i \neq a_j,
\;
0 < (t_j - t_i) \leq \Delta t
\right],
\end{equation}

where \(\mathbf{1}[\cdot]\) denotes the indicator function. Under this definition, the network captures how online search intent transitions into offline visitation behavior, and inversely how physical encounters stimulate subsequent online information seeking activity.

Taking that into account, \textit{Forward Linkage Probabilities} are constructed based on the weighted adjacency matrix 

\begin{equation}
W_u =
\begin{bmatrix}
w_{11} & w_{12} & \cdots & w_{1m} \\
w_{21} & w_{22} & \cdots & w_{2m} \\
\vdots & \vdots & \ddots & \vdots \\
w_{m1} & w_{m2} & \cdots & w_{mm}
\end{bmatrix},
\end{equation}

where \(m = |\mathcal{C}|\) denotes the total number of unique semantic and spatial clusters in the Feedback Network. Given the transition sequence logic, the row represents current cluster while the column is the destination cluster.

Such \textit{Feedback Networks} and \textit{Forward Linkage Probabilities} are calculated for each participant. To estimate population level behavioral dynamics, individual Feedback Networks are aggregated across all individual to construct a Collective Feedback Network:

\begin{equation}
G^{\mathrm{collective}}
=
\left(
V,
\bar{W}
\right),
\end{equation}

where \(V = \bigcup_{u=1}^{N} V_u\) represents the union of all observed clusters across individuals and \(N\) is the number of respondents.

The collective weighted adjacency matrix is computed as the average linkage magnitude across individual networks:

\begin{equation}
\bar{W}
=
\frac{1}{N}
\sum_{u=1}^{N}
W_u,
\end{equation}

such that each element of the aggregate matrix is given by

\begin{equation}
\bar{w}_{pq}
=
\frac{1}{N}
\sum_{u=1}^{N}
w_{pq}^{(u)},
\end{equation}

where \(w_{pq}^{(u)}\) denotes the linkage weight between clusters \(c_p\) and \(c_q\) for individual \(u\). The resulting Collective Feedback Network represents the average cognitive-spatial linkage structure of the urban population, enabling the identification of dominant pathways linking online search behavior and offline visitation activity.

\subsection{Concentration Entropy and Behavioral Push Dynamics}
\label{Concentration Entropy and Behavioral Push Dynamics}
To quantify the degree of concentration or dispersion within the Feedback Network, the \textit{Concentration Entropy} is introduced. The metric evaluates how strongly outgoing linkages from a given cluster converge toward a limited set of destination clusters or diverge across many alternatives. Conceptually, the measure captures whether behavioral dynamics are dominated by routine and repeated preferences or characterized by exploratory and heterogeneous linkages.

Let

\begin{equation}
\mathbf{p}_i
=
\left(
p_{i1}, p_{i2}, \dots, p_{im}
\right)
\end{equation}

denote the row normalized linkage probability vector associated with the origin cluster \(c_i\), where

\begin{equation}
\sum_{j=1}^{m} p_{ij} = 1,
\end{equation}

and \(m = |\mathcal{C}|\) is the total number of destination clusters. The Shannon entropy of cluster \(c_i\), which measures the extent to which the current cluster leads to diverse other clusters linked to it, is computed as

\begin{equation}
H_i
=
-
\sum_{j=1}^{m}
p_{ij}
\log(p_{ij}),
\end{equation}

where terms with \(p_{ij}=0\) are omitted. Because entropy depends on the number of possible clusters connected to the current clusters, the value is normalized by the maximum possible entropy such that

\begin{equation}
H_i^{\mathrm{norm}}
=
\frac{H_i}{\log(m)}.
\end{equation}

To facilitate interpretation in terms of behavioral concentration rather than uncertainty, concentration entropy is defined as the complement of normalized entropy:

\begin{equation}
C_i
=
1 - H_i^{\mathrm{norm}}.
\end{equation}

Higher values of \(C_i\) indicate that outgoing linkages from cluster \(c_i\) are concentrated toward a small number of destinations, implying routine co-occurring behavior, repeated preference structures, but a limited to minimal converging within the cognitive-spatial system. 

Conversely, lower values of \(C_i\) indicate that linkages are distributed across many clusters connected to the current clusters, revealing more co-occurring online and offline behavior and a diverging push across the Feedback Network. Under this interpretation, individuals exhibit diverging cognitive and spatial exploration rather than repeatedly returning to dominant linked behavioral pathways. The interpretation of the concentration entropy index follows the scale presented in Table 1.

\begin{table}[htbp]
\centering
\caption{Average concentration entropy values}
\label{tab:concentration_entropy}
\begin{tabular}{ll}
\hline
\textbf{Concentration Entropy ($C_i$)} & \textbf{Interpretation} \\
\hline
0.0 - 0.3 & Extensive behavioral overlap\\
0.3 - 0.6 & Moderate behavioral overlap\\ 
0.6 - 0.8 & Partial behavioral overlap\\ 
0.8 - 1.0 & Minimal behavioral overlap\\ 
\hline
\end{tabular}
\end{table}
\label{concentration_entropy}

Within the Feedback Network framework, average value of concentration entropy across pair of clusters therefore serves as a compact measure of the strength of co-occurring patterns between online search behavior and offline visitation activity. Low concentration reflects extensive behavioral reinforcement, whereas high concentration reflects the limited exposure stimulating linkages between semantic and spatial domains, with only a small yet dominant number of pairs present.

\section{Results}
\label{Results}
\begin{figure}[!ht]
  \centering
  \includegraphics[width=1.1\textwidth]{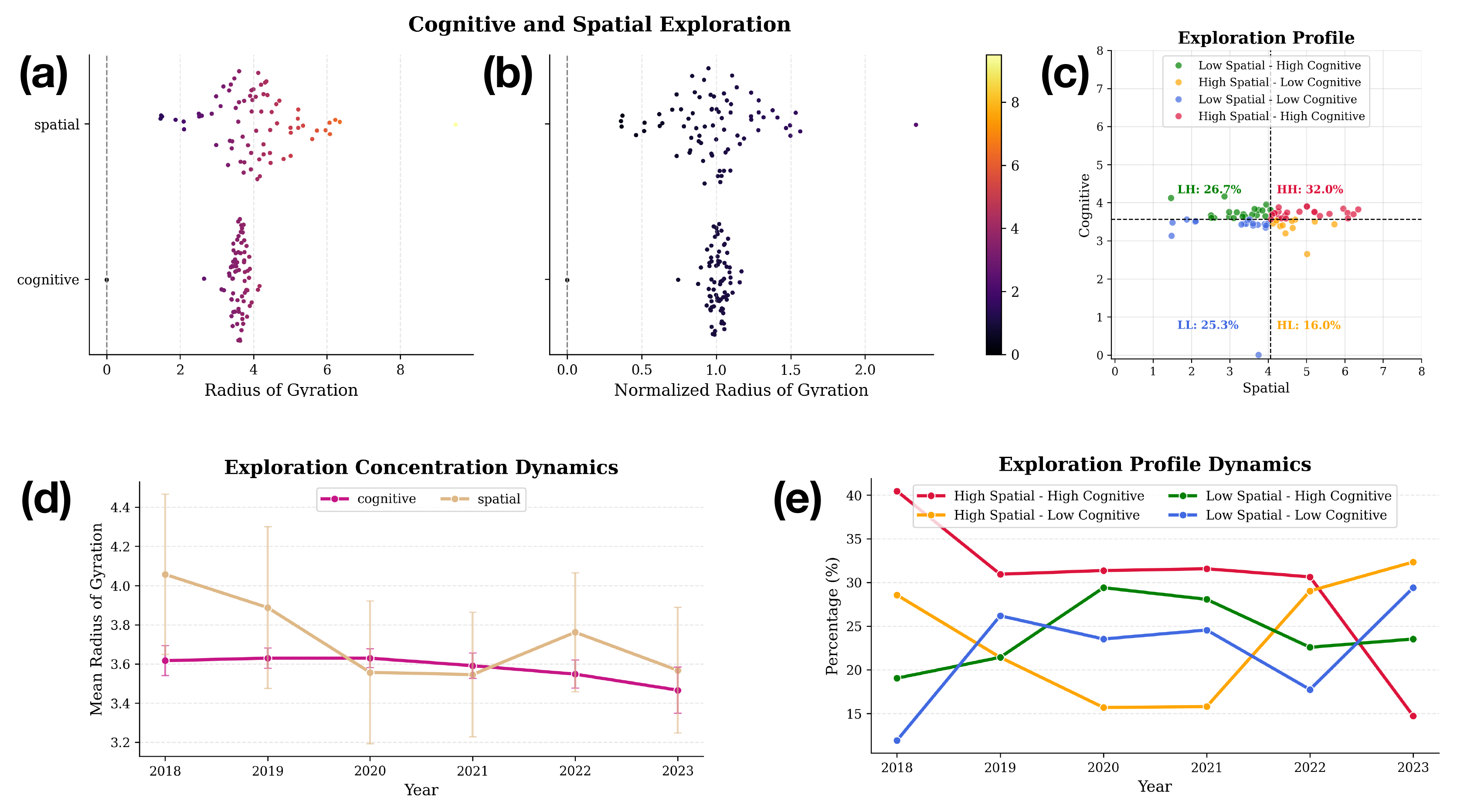}
  \caption{Radius of gyration is computed on the geolocations of visited places (spatial) and on query embeddings (cognitive) (Fig. 3a), and normalized by their mean values (Fig. 3b). Concentration in the online search space is higher than in the offline physical space. Exploration profile (Fig. 3c) with four archetypes based on cognitive and spatial exploration where high alignment in both domains (HH-red) accounts for the largest proportion (32\%). While cognitive  concentration is relatively stable across years, spatial concentration is corrected by the pandemics (2019 to 2020) and rebounds after (Fig. 3d) as supported by drop in HL and increase in LH (Fig. 3e).}
  \label{exploration}
\end{figure}

Describing the characteristics of the online and spatial behaviors, Figure \ref{exploration} shows the overview of exploration measures in both online (cognitive) offline (spatial) space. Cognitive exploration means the extent to which someone varies the topics they search for; while spatial exploration is defined by the extent to which one varies their  locations. Both are measured by the radius of gyration, where low values suggest more routines and concentration and less exploratory patterns, while high values inform about more diverse, exploratory patterns. Based on the distributions of non-normalized (Fig. \ref{exploration}a) and normalized (Fig. \ref{exploration}b) radius of gyration, online search behavior is more concentrated than offline visitation behavior, indicating that individuals tend to reuse a narrower set of query themes than physical destinations, which is relatively more dispersed across space. This pattern is consistent with a stronger pull toward routine and preference reinforcement in the online domain, compared with broader spatial experimentation in the offline domain. The four-archetype exploration profiles (Fig. \ref{exploration}c) further shows that the high high configuration (HH: high spatial, high cognitive) is the largest group in the sample, accounting for 32\% of observations, followed by the low-high (LH: low spatial, high cognitive, 26.7\%), low-low (LL: low spatial, low cognitive, 25.3\%), and high low (HL: high spatial, low cognitive, 16.0\%) types. After all, these results indicate that cognitive and spatial exploration are driven by diverse factors and that they are related but not identical dimensions of behavior, with a substantial share of individuals exhibiting aligned intensity across both domains. 

\clearpage

Across time, cognitive exploration appears comparatively stable from 2018 to 2023, whereas spatial exploration is visibly disrupted around the pandemic period, dropping from 2019 to 2020 and then partially rebounding afterward (Fig. \ref{exploration}d). The profile dynamics are consistent with this shift: the HL share declines after 2019, while LH increases, implying that the pandemic altered the balance between online and offline exploration by constraining physical movement more strongly than digital search behavior (Fig. \ref{exploration}e). These circumstances widened the gap between digital intent and realized mobility before the system gradually recovered. 

\clearpage

\begin{figure}[!ht]
  \centering
  \includegraphics[width=1.1\textwidth]{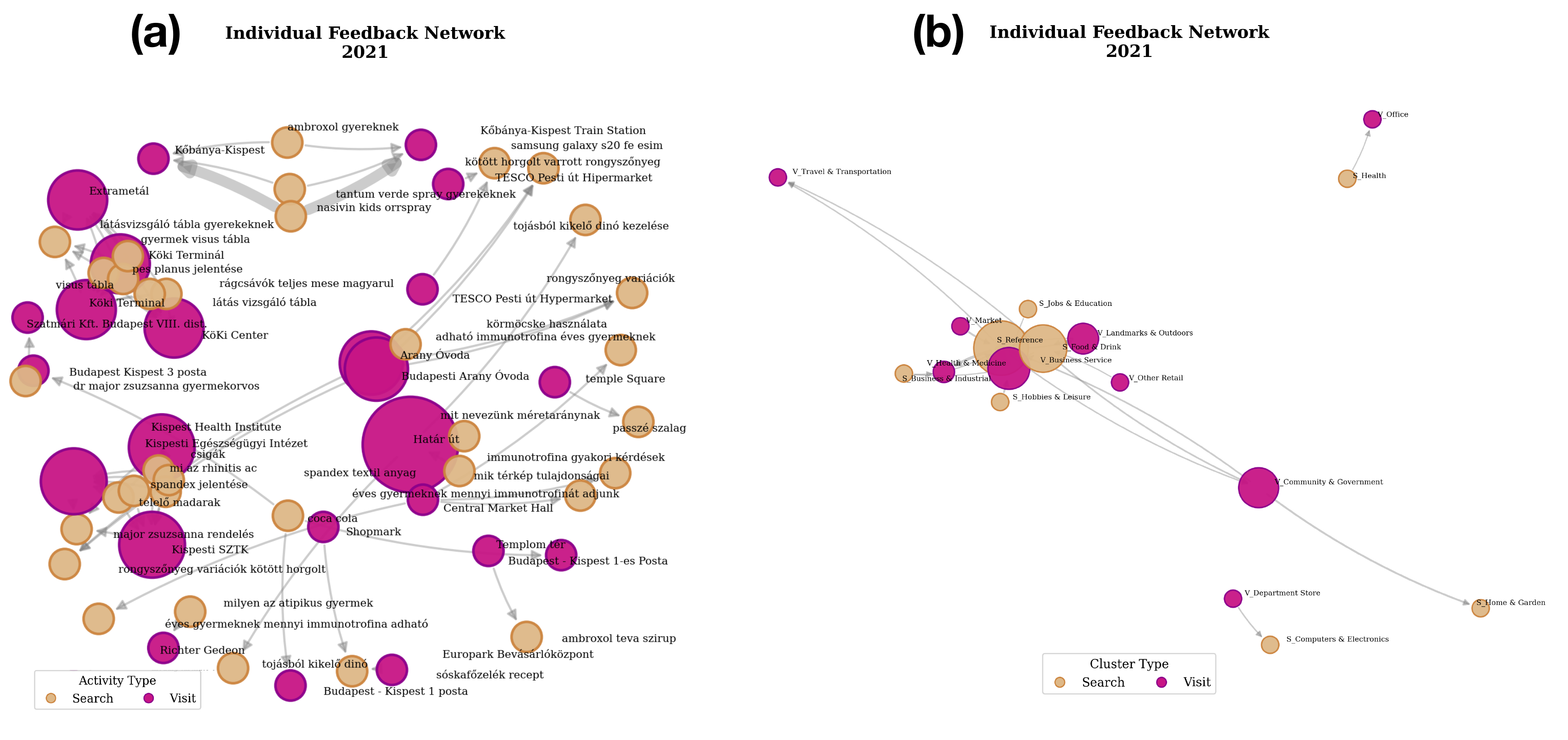}
  \caption{The individual feedback network in 2021 is presented as directed connections at the activity level (Fig. \ref{ind_net}a) and the cluster level (Fig. \ref{ind_net}b), magnifying the interaction between online and offline activities.}  
  \label{ind_net}
\end{figure}

An individual feedback network from 2021 is presented in Figure \ref{ind_net}, illustrating both the original locations and search queries and the same network with the clustered topics. Examining one individual in the dataset, search queries such as ‘ambroxol for children’ (\textit{‘ambroxol gyereknek’}), ‘Tantum Verde spray for children’ (\textit{‘tantum verde spray gyerekeknek’}), and ‘Nasivin Kids nasal spray’ (\textit{‘nasivin kids orrspray’}) lead to a physical presence at a children’s hospital in Kőbánya-Kispest (Fig. \ref{ind_net}a). Considering that this temporal snapshot occurred during the COVID-19 period, the individual appears to have checked health-related conditions or concerns (\textit{S\_Health}) before deciding whether or not to go to the office (\textit{V\_Office}) (Fig. \ref{ind_net}b). 

\clearpage

\begin{figure}[!ht]
  \centering
  \includegraphics[width=1\textwidth]{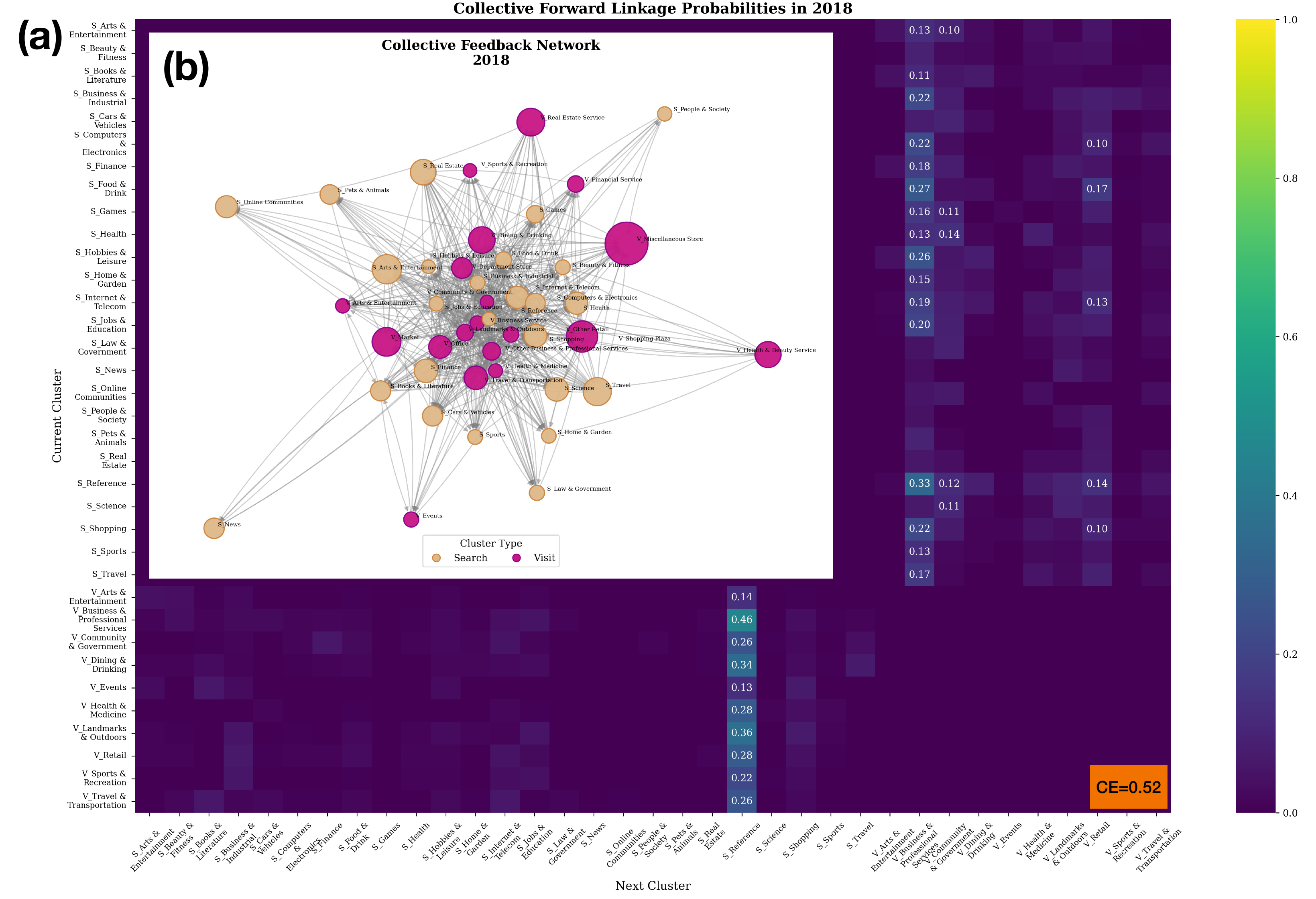}
  \caption{Collective Feedback Network and forward linkage probabilities in 2018. Linkage probability matrix (Fig. \ref{linkage2018}a) reveals heterogeneous cognitive-spatial linkages between online search clusters ('S\_') and offline visit clusters ('V\_'). Higher intensities indicate stronger behavioral linkage probabilities between current (origin) $c_i$ and next (destination) $c_j$ clusters (annotated for $w_{c_i, c_j} \geq 0.1$). The overall concentration entropy ($CE=0.52$) hints at a moderately concentrated linkage structure/overlap characterized by both dominant behavioral pathways and exploratory linkages. Its corresponding Collective Feedback Network (Fig. \ref{linkage2018}b) consists of nodes representing search or visit clusters and weighted edges represent connection intensities between online search and offline visit activities. Node size is scaled by betweenness centrality, reflecting its importance in connecting different clusters.} 
  \label{linkage2018}
\end{figure}

Fig. \ref{linkage2018}b) illustrates the Collective Cognitive-Spatial Feedback Network aggregated across all individuals from the year 2018. In this network those spatial events (location visits) and cognitive events (online searches) are linked, which happened within a one-hour timeframe. The tie goes from the event with the earlier timestamp to the event with the later timestamp. In order to observe robust results, instead of unique events, grouped categories are used: 10 offline visit clusters for locations and 25 search clusters for online queries.

The Collective Feedback Network of that year (Fig. \ref{linkage2018}b) highlights the emergence of highly connected hub clusters linking cognitive (colored in beige) and spatial (colored in magenta) domains where nodes are scaled by betweenness centrality (Supplementary Materials A) to highlight their role as a bridge in the network. Clusters accommodating commercial places such as \textit{V\_Miscellaneous Store}, \textit{V\_Other Retail}, and \textit{V\_Market} highly facilitates behavioral routines and exploration, as indicated by its relatively larger size among visit clusters. Looking at search clusters, queries linked to \textit{S\_Arts \& Entertainment} an\textit{S\_Travel} are more prominent then the rest in coupling digital intention with realized urban activity. At the same time, the network retains multiple cross domain linkages across heterogeneous categories, implying that urban behavioral flows are not strictly compartmentalized but instead exhibit overlapping semantic and spatial interactions. The coexistence of concentrated hubs and peripheral exploratory links is consistent with the intermediate entropy value, reflecting a hybrid structure characterized by both routine reinforcement and exploratory movement in the network.

Based on this network we can calculate the probability that shows how likely it is to search for a topic if one is at a given type of location; and vica versa, how likely it is that one will visit a type of location given they searched for a topic. In case we calculate these measures for the Collective Feedback Network, which aggregates the Feedback Network of all individual we can create the matrix of Collective Forward Linkage Probabilities. This matrix illustrates the probabilities from one physical event to a search event, and reversed.

In 2018, the matrix of Collective Forward Linkage Probabilities (Fig. \ref{linkage2018}a) reveals a mildly concentrated cognitive–spatial linkage structure, with an average concentration entropy value of ($CE=0.52$), indicating that behavioral flows between the physical and the online world are neither fully random nor entirely co-occurring. Instead, the network exhibits a balanced regime in which several dominant pathways coexist with a broader set of lower probability linkages. The heatmap of forward linkage probabilities shows that some clusters generates strong linkage magnitudes as seen in the emergence of prominent linkages pointing from multiple search topics to visit cluster related to \textit{V\_Business \& Professional Services} and from multiple location types to search activity on general information within \textit{S\_Reference} with the highest value $w_{c_i, c_j} = 0.46$. 

Regarding the \textit{S\_Reference} search topic, the types of places from which physical encounters stimulate such subsequent online information seeking behavior are \textit{V\_Arts \& Entertainment}, \textit{V\_Community \& Government}, \textit{V\_Dining \& Drinking}, \textit{V\_Events}, \textit{V\_Health \& Medicine}, \textit{V\_Landmarks \& Outdoor}, \textit{V\_Retail}, \textit{V\_Sports \& Recreation}, and \textit{V\_Travel \& Transportation}. 

On the other hand, strong behavioral conversion from online intention to offline activity in \textit{V\_Business \& Professional Services} is detected from search categories classified under \textit{S\_Reference}, \textit{S\_Shopping}, and \textit{S\_Jobs \& Education}, among others, encouraging physical experiences through visits to business blocks. Moreover, several strong intra-domain co-occurences are also visible, particularly between \textit{S\_Shopping} and \textit{V\_Retail}, reflecting recurrent behavioral routines anchored around consumption oriented urban activities. This indicates that online search behavior and offline visitation activity are organized around behavioral corridors rather than uniformly distributed exploration.

\begin{figure}[!h]
  \centering
  \includegraphics[width=1\textwidth]{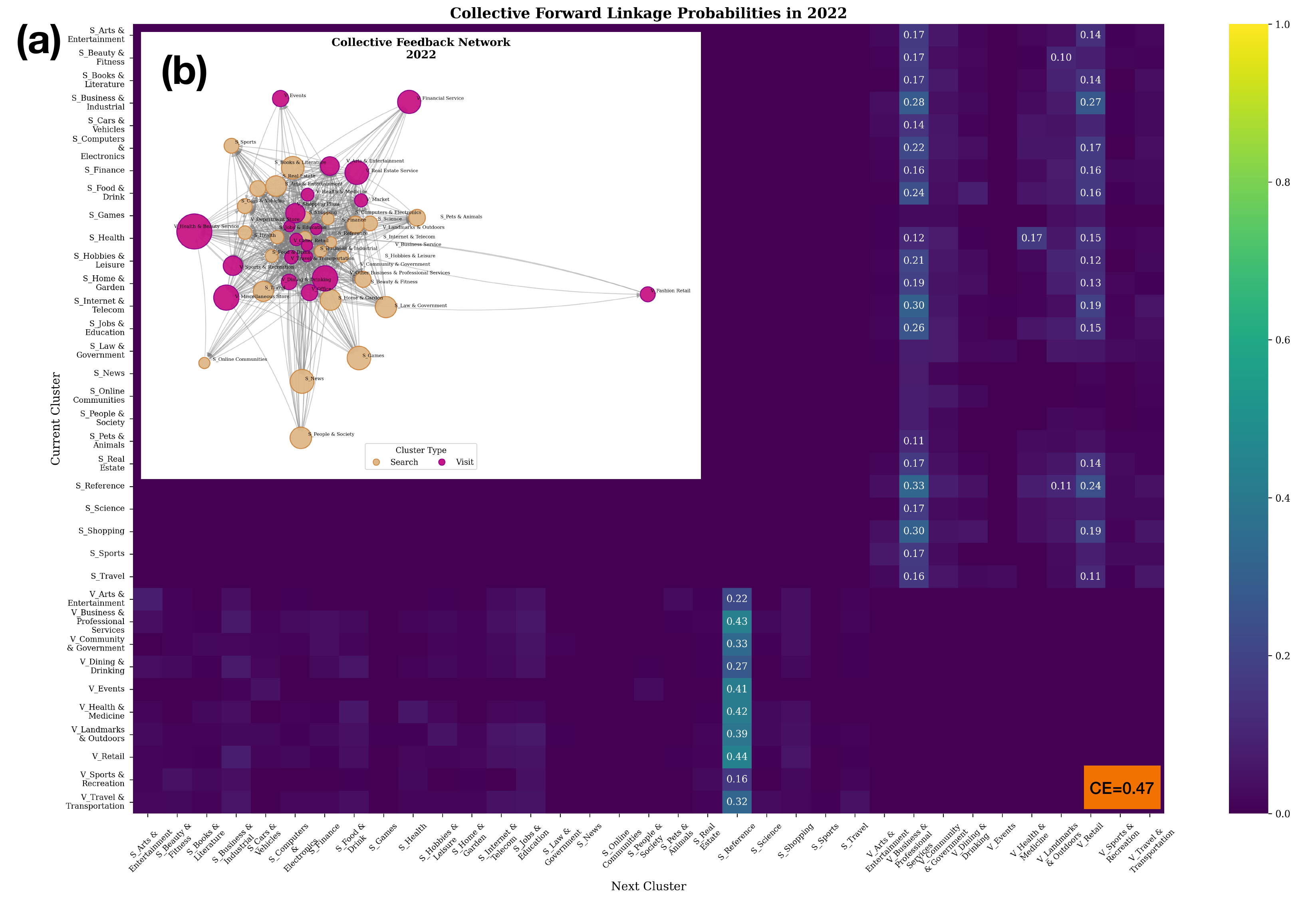}
  \caption{Collective Feedback Network and cognitive-spatial linkage structure in 2022. The linkage matrix (Fig. \ref{linkage2022}a) illustrates forward linkage probabilities between online search categories (“S\_”) and offline visitation categories (“V\_”) and vica vera, where brighter cells correspond to stronger linkage likelihoods from origin cluster $c_i$ to destination cluster $c_j$ (displayed for $w_{c_i, c_j} \geq 0.1$). Compared with earlier years the figure indicates more significant co-occuring patterns between the online and the offline events, implying that linkages become less dominated by a small set of recurrent pathways. Strong cognitive-spatial corridors can be observed among retail, dining, lifestyle, and entertainment related activities, reflecting continued reinforcement between digital intention and physical urban engagement. The corresponding network representation (Fig. \ref{linkage2022}b) visualizes the Collective Feedback Network, where nodes denote semantic search clusters or spatial visitation clusters and weighted edges indicate linkage strengths across domains. Node size is proportional to betweenness centrality (Supplementary Materials A), highlighting clusters that function as key intermediaries linking multiple behavioral pathways within the urban system.}  
  \label{linkage2022}
\end{figure}

The Collective Feedback Network from 2022 (Fig. \ref{linkage2022}b) illustrates the evolving structure of cognitive-spatial interactions in the last analyzed year. Nodes representing spatial visitation clusters (magenta) and semantic search clusters (beige) remain highly interconnected, but the network topology appears more distributed than in 2018. Commercially oriented visit clusters such as \textit{V\_Miscellaneous Store} continue to function as important bridging hubs with relatively high betweenness centrality. Another change detected in this post-pandemic year is the strengthening position of health conscious visit cluster embodied in \textit{V\_Health \& Beauty Services}, indicating its elevating role in facilitating linkages between multiple cognitive and spatial domains. In parallel, search clusters linked to amusement and information such as \textit{S\_News}, \textit{S\_People \& Society}, \textit{S\_Games}, and \textit{S\_Books \& Literature} become increasingly prominent in connecting heterogeneous pathways across the network. In contrast to the configuration in 2018, where dominant hubs produce a more concentrated hybrid structure, the 2022 network exhibits a more decentralized arrangement characterized by a larger number of intermediate strength linkages.

The calculation of the matrix of Collective Forward Linkage Probabilities from the Collective Feedback Network in 2022 (Fig. \ref{linkage2022}a) exhibits a slightly more dispersed cognitive-spatial linkage structure compared with 2018. This demonstrates that significant co-occurring behavioral linkages between the online and the offline space became a bit more dominant. At the same time, there are multiple spatial and online clusters that do not show strong co-occurring patterns at all. 

Focusing on links from spatial to cognitive events, the forward linkage probability matrix reveals that the strongest linkage pathway emerges from \textit{V\_Retail} to \textit{S\_Reference}, with a linkage probability of $w_{c_i,c_j}=0.44$, higher than the corresponding linkage observed in 2018 but slightly lower than the strongest linkage from \textit{V\_Business \& Professional Services} to \textit{S\_Reference} in 2018. The observed patterns also persist across multiple visitation clusters, including \textit{V\_Dining \& Drinking}, \textit{V\_Retail}, \textit{V\_Health \& Medicine}, \textit{V\_Landmarks \& Outdoors}, and \textit{V\_Travel \& Transportation}, indicating that physical encounters continue to stimulate subsequent online information seeking activity on (\textit{S\_Reference}).

The reverse direction from online search intention to offline activity also remains evident in 2022. Search categories associated with \textit{S\_Reference}, \textit{S\_Shopping}, and \textit{S\_People \& Society} continue to channel users toward visitation clusters related to commercial and service oriented destinations (e.g.: \textit{V\_Business \& Professional Services} and \textit{V\_Retail}). Interestingly, strong semantic-spatial coupling remains visible between \textit{S\_Shopping} and \textit{V\_Retail}, indicating that consumption related behavioral routines still represent one of the most stable pathways in the Feedback Network. Nevertheless, compared with 2018, the linkage landscape appears less centralized around consumption and professional-service activities alone, with a larger number of moderate-strength linkages emerging across entertainment, health, recreation, and mobility related categories. This broader distribution of linkages suggests that cognitive and spatial activities became behaviorally more diversified in 2022.

\begin{figure}[!ht]
  \centering
  \includegraphics[width=1.1\textwidth]{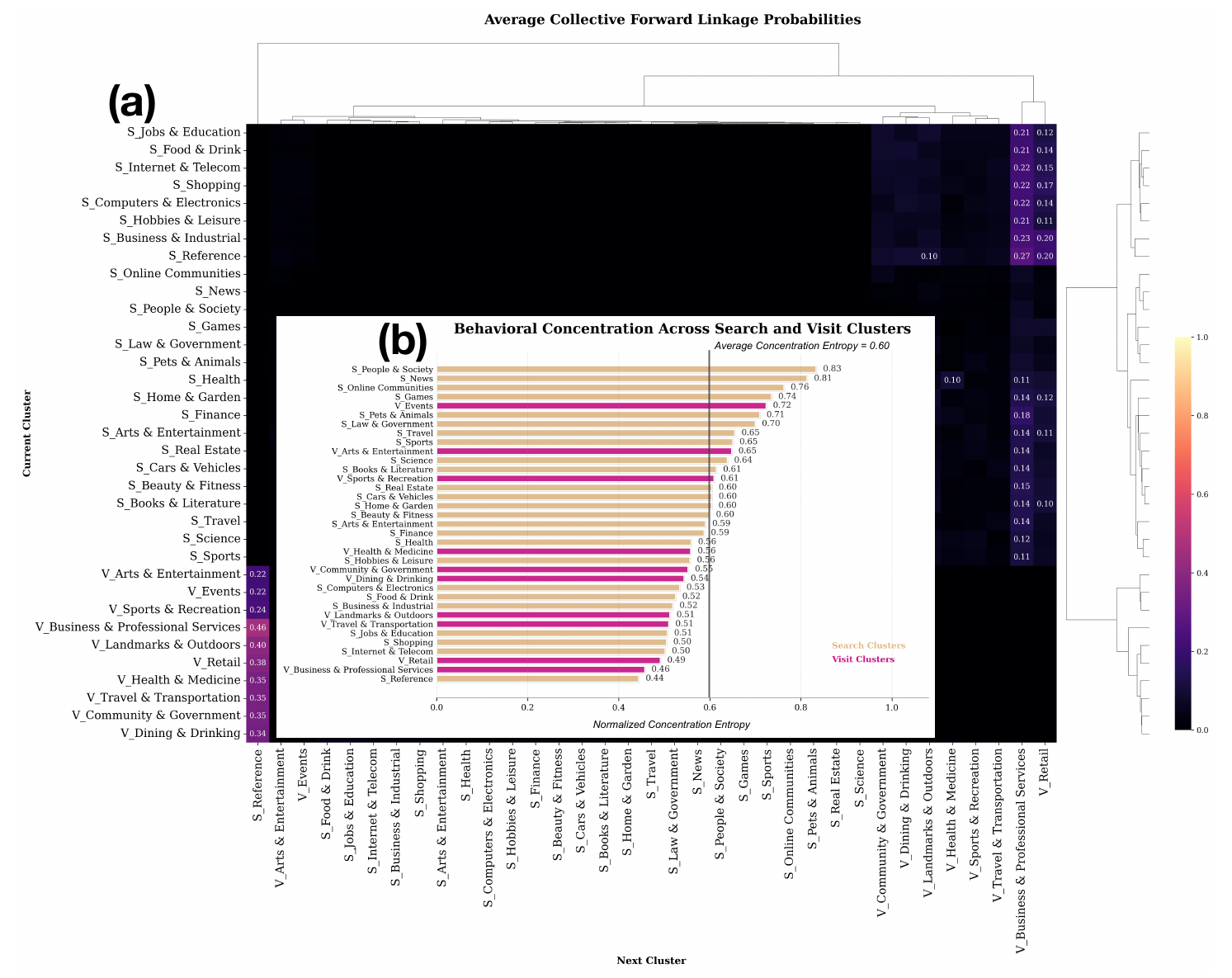}
  \caption{The average forward linkage matrix (Fig. \ref{avg_linkage}a) summarizes heterogeneous linkage probabilities between online search categories ('S\_') and offline visitation categories ('V\_') over the entire observation period (2018-2022). Cell value represents the mean probability of transitioning from origin cluster $c_i$ to destination cluster $c_j$ with annotations displayed for stronger linkages ($w_{c_1, c_2} \geq 0.1$). Hierarchical dendogram on the right (Supplementary Materials B) reveals a visible split between search (upper rows) and visit (lower rows) cluster at coarse grain level. Persistent corridors are notably visible between \textit{V\_Business \& Professional Services}, and \textit{S\_Reference}. Behavioral concentration across clusters are summarized in Fig. \ref{avg_linkage}b, specifying normalized concentration entropy by row (current cluster) where lower values indicate the existence of more linkages. The average concentration entropy ($CE=0.60$) indicates a moderate level of behavioral overlap.}  
  \label{avg_linkage}
\end{figure}

Figure \ref{avg_linkage}a presents the Average Collective Feedback Network across the whole 2018-2022 observation period and reveals the emergence of a persistent long term cognitive-spatial interaction structure linking online search behavior with offline visitation activity. It demonstrates that behavioral linkages are not uniformly distributed across semantic and spatial categories. Contrastingly, a we can detect co-occurring pathways in the overall network dynamics. Strong linkage intensities remain concentrated around commercial, consumption oriented, and informational activities, indicating that certain forms of online intention consistently translate into recurring physical mobility patterns over time and vice versa.

In particular, the strongest linkage is observed between \textit{V\_Business \& Professional Services} and \textit{S\_Reference} ($w_{c_i, c_j} = 0.46$). It is worth highlighting that mostly people end up with general information retrieval activities under cluster \textit{S\_Reference} as a follow up action after physical exposure through visit to different locations such as in \textit{V\_Landmarks\& Outdoors}, \textit{V\_Retails}, and \textit{V\_Travel \& Transportation}. Therefore, such patterns infer that encounters with business and experience oriented urban spaces repeatedly stimulate subsequent online information seeking behavior. In  addition, there are similar high probability destination clusters, for instance \textit{V\_Business \& Professional Services} and \textit{V\_Retail}, reflecting robust semantic-spatial reinforcement loops embedded within everyday urban routines. 

\clearpage

The hierarchical configuration structure further indicates that several semantic and spatial categories form coherent behavioral communities. Search clusters associated with shopping, food, internet services, entertainment, and hobbies are grouped closely with visitation clusters related to retail, dining, landmarks, and recreational activities. This clustering pattern implies that online cognitive exposure and offline spatial encounters are systematically interconnected rather than operating independently. In other words, urban behavioral flows exhibit functional specialization where related semantic interests repeatedly correspond to specific categories of physical destinations. Concurrently, the presence of weaker but widespread peripheral linkages implies that exploratory pathways remain embedded within the system, even though they contribute less strongly to the overall network structure.

The concentration profile shown in Fig.~\ref{avg_linkage}b reveals substantial variation in behavioral concentration across current clusters. The values in each bar represent the normalized concentration entropy for each row (current cluster). Because lower values indicate the presence of more linkages, \textit{S\_Reference} exhibits substantially more connections between current and subsequent clusters. It is reasonable to emphasize that individuals tend to be exposed to general information before taking further actions or making decisions such as visiting places in cluster \textit{S\_Landmarks \& Outdoors}, \textit{S\_Business \& Professional Services}, and \textit{S\_Retail}, while also seeking additional information following a particular physical encounter across various locations.

The average concentration entropy value ($CE=0.60$) indicates a moderately high degree of concentration, implying that long term behavioral dynamics are shaped more strongly by recurrent routines than by uniformly distributed exploration. Search related clusters generally exhibit higher normalized concentration entropy than visitation clusters, indicating that online search behavior is semantically narrower and more repetitive than offline mobility behavior. Categories such as \textit{S\_People \& Society}, \textit{S\_News}, \textit{S\_Online Communities}, and \textit{S\_Games} display  high values, indicating that users repeatedly return to a limited set of semantic interests within these domains. In contrast, visitation clusters such as \textit{V\_Travel \& Transportation}, \textit{V\_Landmarks \& Outdoors}, and \textit{V\_Retail} exhibit comparatively lower values, reflecting broader and more exploratory patterns of spatial engagement across urban environments. While coupling between online and offline activities becomes evident and more prominently observable at higher resolution through the Individual Feedback Networks (Fig. \ref{ind_net}), aggregated Collective Feedback Networks still show strong, but very generally interpretable co-occurrences of spatial and online behavior.

\clearpage

\section{Discussion}
\label{Discussion}
This paper contributes to the knowledge of digital twins by analyzing how the relationship between offline mobility and online search diversity evolves over time, revealing co-occurrences in cognitive-spatial engagement. To explore the conditions shaping these dynamics, we further analyze the search keywords using NLP and LLM, modeling search histories as time stamped corpora and extracting their semantic structure to characterize the cognitive dimension of online exploration. To do so, contextual embedding model is employed to map queries into a high dimensional semantic space, allowing us to measure semantic dispersion and thematic drifts across years. Linking these semantic measures to identification of activities in the offline visit space through reverse geocoding allows us to assess whether physical mobility to certain types of places is accompanied by thematic concentration or diversification in online behavior. The methodology and the model helps clarifying how online and offline spatial behavior are co-occurring through the conceptual framework of the Feedback Network and Concentration Entropy.

The findings reveal that online search behavior and offline visitation activity are interconnected yet structurally asymmetric dimensions of urban mobility. In the radius of gyration analysis, individuals consistently exhibit higher concentration in cognitive online exploration than in spatial exploration, indicating that digital attention tends to converge around narrower semantic interests while physical mobility remains comparatively more diverse. This asymmetry signals the presence of informational exposure in digital environments is more strongly shaped by routine reinforcement and repeated preference structures than movement through urban space. Such a pattern aligns with broader concerns regarding algorithmic filtering, recommendation systems, and selective exposure mechanisms that increasingly channel users toward recurrent informational pathways rather than heterogeneous discovery. In the meantime, the persistence of broader spatial exploration indicates that offline urban environments continue to provide opportunities for diverse encounters that are not fully reducible to online behavioral routines.

The Feedback Network demonstrates that these two domains are not isolated systems but instead form a cognitive-spatial loop. Strong linkages between semantic search categories and physical visitation clusters demonstrate that online intention and offline behavior mutually reinforce one another through repeated behavioral corridors. Stable pathways emerge around consumption oriented, professional service, mobility, and entertainment related online and spatial activities, indicating that many urban routines are increasingly mediated through digital information ecosystems, which is highly relevant to the operational design of Digital Twins in urban social dynamics (Hybrid Digital Twin of Society). The coexistence of concentrated hubs and weaker transitions between online and spatial events reveals that urban behavior simultaneously contains both converging and diverging dynamics where some activities become stabilized into recurrent cognitive-spatial links, while others continue to facilitate exploratory  or diverging behavior. Simultaneously, this recurring pattern also reflects the urban morphology and spatial fabric of Budapest, where spatial signatures associated with business and professional services (e.g.: commercial districts) and retail environments (e.g.: malls and shopping plazas) strongly intersect with everyday mobility patterns.

Temporal comparisons also highlight the sensitivity of this relationship to external disruption. The pandemic period reduced spatial concentration while leaving cognitive concentration relatively stable, implying that digital exploration became partially decoupled from realized mobility under conditions of constrained movement. From this perspective, it also reflects the impact of digitalization in Hungary, where the data used in this paper were collected, as an increasing number of everyday errands can now be performed through online platforms, with the pandemic serving as a catalyst for the wave of digitalization. Consequently, the distance between online and offline activities continues to diminish. The subsequent rebound in spatial concentration during the post-pandemic period pinpoints the gradual recovery of physical routines, with the overall Feedback Network in later years showing more significant co-occurrences between the online and the physical events compared to 2018. This indicates that large scale disruptions may permanently reshape the balance between digital exposure and physical encounter by altering how online information substitutes for or complements urban movement.

More broadly, the proposed framework contributes to the emerging literature on behavioral Digital Twins by extending urban modeling beyond infrastructure and trajectory analysis toward the integration of cognitive digital activity and semantic behavior. Existing urban Digital Twins typically represent cities through physical systems such as transportation networks, buildings, and environmental sensors. By contrast, the Feedback Network conceptualizes cities as hybrid socio-digital systems in which online attention, informational demand, and physical mobility continuously interact. Within this perspective, Concentration Entropy provides a useful mechanism for quantifying the degree to which urban behavior is driven by convergent routine structures versus divergent exploratory dynamics. The framework therefore offers a foundation for future research examining cognitive-spatial linkage, digital mediation of urban mobility, and the role of information ecosystems in shaping patterns of urban co-presence.

A limitation of this study concerns the sample size (N=90) of representative Budapest residents. Obtaining both location and search data from the same participants is relatively rare. Despite this limitation and taking the long temporal range of the data into account, the existing dataset is sufficient for a proof-of-concept analysis, particularly because interpretable structures become even more apparent at the level of individual feedback networks. These patterns could be examined in greater depth using larger-scale datasets. Therefore, future research may extend this approach by incorporating larger datasets beyond the existing data donation framework and expanding observations across multiple cities and countries to validate the robustness of linkages between online and offline activities. While the use of standardized clustering categories based on Google Trends (search clusters) and Foursquare (visit clusters) benchmarks is justified and methodologically reasonable, there remains substantial room to improve cluster granularity by subdividing broad categories such as \textit{S\_Reference}, \textit{V\_Business \& Professional Services}, and \textit{V\_Retail}. Such refinement becomes particularly important when advancing toward more socially interpretable causal modeling of cognitive-spatial behavioral dynamics where mobility, accessibility, and linkage in digitally mediated urban environments are continuously shaped and reshaped. \\

\section{Conclusion}
\label{Conclusion}
This paper introduces the concept of the Feedback Network as a computational framework for linking online search behavior and offline visitation activity within a unified cognitive-spatial system. By integrating semantic search exploration with physical mobility trajectories, this paper demonstrates that urban behavior increasingly unfolds across interconnected digital and spatial environments rather than within purely physical space alone. The proposed framework operationalizes this relationship through heterogeneous linkage networks connecting online search clusters and offline visitation clusters, while Concentration Entropy was developed to quantify the balance between routine reinforcement and exploratory behavior across the system.

The empirical findings directly address the two research questions posed in this paper. Regarding the first question on whether individuals behave with different diversity when exploring offline visit spaces versus online search spaces, the results demonstrate a clear asymmetry between cognitive and spatial exploration. Cognitive exploration consistently exhibits higher concentration than spatial exploration, indicating that online behavior is more strongly shaped by routine reinforcement and closely related semantic interests, whereas offline mobility remains comparatively more diverse and exploratory due to uneven spatial distribution of locations and attribution towards distance decay effect in urban configuration. This divergence shows that digital environments increasingly channel attention through recurrent informational pathways, while physical urban space continues to preserve broader opportunities for heterogeneous encounters and movement. Moreover, temporal disruptions such as the COVID-19 pandemic show that the strength and concentration of these behavioral links between spatial and online events are dynamic and sensitive to external shocks, particularly through the temporary weakening of spatial routines while cognitive concentration remains relatively stable.

Concerning the second research question on how robust behavioral links can be modeled through a cross domain Feedback Network linking online search activity and offline visitation behavior, the analysis reveals the existence of stable cognitive-spatial corridors connecting semantic search clusters with physical destination clusters. Persistent linkages between commercial, informational, entertainment, and mobility related categories demonstrate that online intention and offline behavior form structures rather than isolated domains. The proposed Feedback Network successfully captures these linkages by representing how search behavior stimulates physical movement and how physical encounters subsequently generate new cycles of online information seeking. 

In summary, based on these findings, it is reasonably argued that contemporary urban systems can no longer be understood solely through trajectories of physical movement. Instead, mobility increasingly reflects the interaction between informational exposure and spatial encounter, where online attention shapes physical accessibility and urban experiences simultaneously generate new cycles of digital engagement. The Feedback Network framework therefore provides a conceptual and computational foundation for future Hybrid Digital Twin of Society capable of modeling not only where people move, but also how cognitive attention, semantic exploration, and informational ecosystems influence urban dynamics. \\

\textbf{CRediT authorship contribution statement}\\
\textbf{Rafiazka Hilman}: Conceptualization, Methodology, Software, Validation, Formal analysis, Investigation, Resources, Data Curation, Writing – Original
Draft, Writing – Review \& Editing,  Visualization, Funding acquisition\\ 
\textbf{Júlia Koltai}: Conceptualization, Validation, Resources, Writing – Review \& Editing, Supervision, Project administration, Funding acquisition\\

\textbf{Declaration of competing interest}\\
The authors declare that they have no known competing financial interests or personal relationships that could have appeared to influence the work reported in this paper.\\

\textbf{Funding sources}\\
Rafiazka Hilman acknowledges support from the Momentum MSCA Programme co-funded by the European Commission through the HORIZON-MSCA-2023-COFUND programme and the Secretariat of the Hungarian Academy of Sciences (MTA) under Grant Agreement No. 101179854. Julia Koltai acknowledges funding from the
Hungarian Academy of Sciences Lendület Program: LP2022-10/2022. \\

\textbf{Declaration of generative AI and AI-assisted technologies in the manuscript preparation process}\\
During the preparation of this work, the author(s) used GPT-5.5 (the latest flagship model of ChatGPT) for pseudo-validation/approximate validation of clustering results for the Google Trends and Foursquare categories used in the pipeline. After using this tool/service, the author(s) reviewed and edited the content as needed and take(s) full responsibility for the content of the published article.\\

\clearpage

\newpage

\bibliography{sample}

@article{breuer2023user,
  title={User-centric approaches for collecting Facebook data in the ‘post-API age’: Experiences from two studies and recommendations for future research},
  author={Breuer, Johannes and Kmetty, Zolt{\'a}n and Haim, Mario and Stier, Sebastian},
  journal={Information, Communication \& Society},
  volume={26},
  number={14},
  pages={2649--2668},
  year={2023},
  publisher={Taylor \& Francis}
}

@article{gonzalez2008understanding,
  title={Understanding individual human mobility patterns},
  author={Gonzalez, Marta C and Hidalgo, Cesar A and Barabasi, Albert-Laszlo},
  journal={nature},
  volume={453},
  number={7196},
  pages={779--782},
  year={2008},
  publisher={Nature Publishing Group UK London}
}

@misc{kmetty_donacio_2024,
	type = {Data {Collection}},
	title = {Donáció alapú digitális adatgyűjtés},
	copyright = {cc\_by\_nd},
	url = {https://openarchive.tk.mta.hu/629/},
	language = {hu},
	urldate = {2026-05-18},
	author = {Kmetty, Zoltán and Koltai, Júlia and Stefkovics, Ádám and Rakovics, Zsófia and Knap, Árpád and Váradi, Bendegúz},
	month = nov,
	year = {2024},
}

@inproceedings{kossinets2008structure,
  title={The structure of information pathways in a social communication network},
  author={Kossinets, Gueorgi and Kleinberg, Jon and Watts, Duncan},
  booktitle={Proceedings of the 14th ACM SIGKDD international conference on Knowledge discovery and data mining},
  pages={435--443},
  year={2008}
}

@article{louail2014mobile,
  title={From mobile phone data to the spatial structure of cities},
  author={Louail, Thomas and Lenormand, Maxime and Cantu Ros, Oliva G and Picornell, Miguel and Herranz, Ricardo and Frias-Martinez, Enrique and Ramasco, Jos{\'e} J and Barthelemy, Marc},
  journal={Scientific reports},
  volume={4},
  number={1},
  pages={5276},
  year={2014},
  publisher={Nature Publishing Group UK London}
}

@article{song2010limits,
  title={Limits of predictability in human mobility},
  author={Song, Chaoming and Qu, Zehui and Blumm, Nicholas and Barab{\'a}si, Albert-L{\'a}szl{\'o}},
  journal={Science},
  volume={327},
  number={5968},
  pages={1018--1021},
  year={2010},
  publisher={American Association for the Advancement of Science}
}

@book{manning2008introduction,
  title={Introduction to information retrieval},
  author={Manning, Christopher D},
  year={2008},
  publisher={Syngress Publishing,}
}

@inproceedings{reimers2019sentence,
  title={Sentence-bert: Sentence embeddings using siamese bert-networks},
  author={Reimers, Nils and Gurevych, Iryna},
  booktitle={Proceedings of the 2019 conference on empirical methods in natural language processing and the 9th international joint conference on natural language processing (EMNLP-IJCNLP)},
  pages={3982--3992},
  year={2019}
}

@article{wang2022text,
  title={Text embeddings by weakly-supervised contrastive pre-training},
  author={Wang, Liang and Yang, Nan and Huang, Xiaolong and Jiao, Binxing and Yang, Linjun and Jiang, Daxin and Majumder, Rangan and Wei, Furu},
  journal={arXiv preprint arXiv:2212.03533},
  year={2022}
}

@article{mcinnes2018umap,
  title={Umap: Uniform manifold approximation and projection for dimension reduction},
  author={McInnes, Leland and Healy, John and Melville, James},
  journal={arXiv preprint arXiv:1802.03426},
  year={2018}
}

@inproceedings{campello2013density,
  title={Density-based clustering based on hierarchical density estimates},
  author={Campello, Ricardo JGB and Moulavi, Davoud and Sander, J{\"o}rg},
  booktitle={Pacific-Asia conference on knowledge discovery and data mining},
  pages={160--172},
  year={2013},
  organization={Springer}
}

@article{stopher2007assessing,
  title={Assessing the accuracy of the Sydney Household Travel Survey with GPS},
  author={Stopher, Peter and FitzGerald, Camden and Xu, Min},
  journal={Transportation},
  volume={34},
  number={6},
  pages={723--741},
  year={2007},
  publisher={Springer}
}

@article{barbosa2018human,
  title={Human mobility: Models and applications},
  author={Barbosa, Hugo and Barthelemy, Marc and Ghoshal, Gourab and James, Charlotte R and Lenormand, Maxime and Louail, Thomas and Menezes, Ronaldo and Ramasco, Jos{\'e} J and Simini, Filippo and Tomasini, Marcello},
  journal={Physics Reports},
  volume={734},
  pages={1--74},
  year={2018},
  publisher={Elsevier}
}

@misc{batty2018digital,
  title={Digital twins},
  author={Batty, Michael},
  journal={Environment and planning B: Urban analytics and City Science},
  volume={45},
  number={5},
  pages={817--820},
  year={2018},
  publisher={Sage Publications Sage UK: London, England}
}

@article{dembski2020urban,
  title={Urban digital twins for smart cities and citizens: The case study of Herrenberg, Germany},
  author={Dembski, Fabian and W{\"o}ssner, Uwe and Letzgus, Mike and Ruddat, Michael and Yamu, Claudia},
  journal={Sustainability},
  volume={12},
  number={6},
  pages={2307},
  year={2020},
  publisher={MDPI}
}

@article{lazer2009social,
  title={Social science. Computational social science.},
  author={Lazer, David and Pentland, Alex and Adamic, Lada and Aral, Sinan and Barabasi, Albert-Laszlo and Brewer, Devon and Christakis, Nicholas and Contractor, Noshir and Fowler, James and Gutmann, Myron and others},
  journal={Science (New York, NY)},
  volume={323},
  number={5915},
  pages={721--723},
  year={2009}
}

@article{alessandretti2020scales,
  title={The scales of human mobility},
  author={Alessandretti, Laura and Aslak, Ulf and Lehmann, Sune},
  journal={Nature},
  volume={587},
  number={7834},
  pages={402--407},
  year={2020},
  publisher={Nature Publishing Group UK London}
}

@article{pappalardo2015returners,
  title={Returners and explorers dichotomy in human mobility},
  author={Pappalardo, Luca and Simini, Filippo and Rinzivillo, Salvatore and Pedreschi, Dino and Giannotti, Fosca and Barab{\'a}si, Albert-L{\'a}szl{\'o}},
  journal={Nature communications},
  volume={6},
  number={1},
  pages={8166},
  year={2015},
  publisher={Nature Publishing Group UK London}
}

@article{kitchin2014real,
  title={The real-time city? Big data and smart urbanism},
  author={Kitchin, Rob},
  journal={GeoJournal},
  volume={79},
  number={1},
  pages={1--14},
  year={2014},
  publisher={Springer}
}

@article{shelton2015actually,
  title={The ‘actually existing smart city’},
  author={Shelton, Taylor and Zook, Matthew and Wiig, Alan},
  journal={Cambridge journal of regions, economy and society},
  volume={8},
  number={1},
  pages={13--25},
  year={2015},
  publisher={Oxford University Press UK}
}

@article{verhoef2015multi,
  title={From multi-channel retailing to omni-channel retailing: introduction to the special issue on multi-channel retailing},
  author={Verhoef, Peter C and Kannan, Pallassana K and Inman, J Jeffrey},
  journal={Journal of retailing},
  volume={91},
  number={2},
  pages={174--181},
  year={2015},
  publisher={Elsevier}
}

@article{huang2020twitter,
  title={Twitter reveals human mobility dynamics during the COVID-19 pandemic},
  author={Huang, Xiao and Li, Zhenlong and Jiang, Yuqin and Li, Xiaoming and Porter, Dwayne},
  journal={PloS one},
  volume={15},
  number={11},
  pages={e0241957},
  year={2020},
  publisher={Public Library of Science San Francisco, CA USA}
}

@misc{googleGoogleTrends,
	author = {Google},
	title = {{G}oogle {T}rends --- trends.google.com},
	howpublished = {\url{https://trends.google.com/trends/?hl=en-{U}{S}}},
	year = {2026},
	note = {[Accessed 26-05-2026]},
}

@misc{foursquareFoursquareCategories,
	author = {Foursquare},
	title = {{F}oursquare {C}ategories and {C}ore {A}ttributes | {F}oursquare --- docs.foursquare.com},
	howpublished = {\url{https://docs.foursquare.com/data-products/docs/categories}},
	year = {2016},
	note = {[Accessed 26-05-2026]},
}

@article{yossef2023social,
  title={The social digital twin: The social turn in the field of smart cities},
  author={Yossef Ravid, Batel and Aharon-Gutman, Meirav},
  journal={Environment and Planning B: Urban Analytics and City Science},
  volume={50},
  number={6},
  pages={1455--1470},
  year={2023},
  publisher={SAGE Publications Sage UK: London, England}
}

@inproceedings{chircu2023towards,
  title={Towards a digital twin of society},
  author={Czarnecki, Christian and Friedmann, Daniel and Pomaskow, Joanna and Sultanow, Eldar},
  booktitle={Proceedings of the 56th Hawaii International Conference on System Sciences},
  year={2023}
}

@article{zinman2020utilizing,
  title={Utilizing digital traces of mobile phones for understanding social dynamics in urban areas},
  author={Zinman, Oded and Lerner, Boaz},
  journal={Personal and Ubiquitous Computing},
  volume={24},
  number={4},
  pages={535--549},
  year={2020},
  publisher={Springer}
}

@inproceedings{vaccari2009holistic,
  title={A holistic framework for the study of urban traces and the profiling of urban processes and dynamics},
  author={Vaccari, Andrea and Liu, Liang and Biderman, Assaf and Ratti, Carlo and Pereira, Francisco and Oliveirinha, Joao and Gerber, Alexandre},
  booktitle={2009 12th international IEEE conference on intelligent transportation systems},
  pages={1--6},
  year={2009},
  organization={IEEE}
}

@article{ferre2022adoption,
  title={The adoption of urban digital twins},
  author={Ferr{\'e}-Bigorra, Jaume and Casals, Miquel and Gangolells, Marta},
  journal={Cities},
  volume={131},
  pages={103905},
  year={2022},
  publisher={Elsevier}
}

@inproceedings{brelsford2019using,
  title={Using digital trace data to identify regions and cities},
  author={Brelsford, Christa and Thakur, Gautam and Arthur, Rudy and Williams, Hywel},
  booktitle={Proceedings of the 2nd ACM SIGSPATIAL International Workshop on Advances on Resilient and Intelligent Cities},
  pages={5--8},
  year={2019}
}

@article{zhou2026exponential,
  title={Exponential distance decay in urban park visitation: A comparative analysis of recreational mobility across 20 US metropolitan areas},
  author={Zhou, Yichun and Huang, Xiao and Li, Zhenlong and Guo, Qianwen Vivian and Wei, Hanxue},
  journal={Landscape and Urban Planning},
  volume={273},
  pages={105690},
  year={2026},
  publisher={Elsevier}
}

@article{ceder2021urban,
  title={Urban mobility and public transport: future perspectives and review},
  author={Ceder, Avishai},
  journal={International Journal of Urban Sciences},
  volume={25},
  number={4},
  pages={455--479},
  year={2021},
  publisher={Taylor \& Francis}
}

@article{elorduy2025assessing,
  title={Assessing public transport accessibility for people with physical disabilities in burgos, spain: A user-centered approach to inclusive urban mobility},
  author={Elorduy, Juan L and Pino, Yesica and Gento, {\'A}ngel M},
  journal={Plos one},
  volume={20},
  number={4},
  pages={e0322068},
  year={2025},
  publisher={Public Library of Science San Francisco, CA USA}
}

@article{wang2022land,
  title={Land use spatial optimization using accessibility maps to integrate land use and transport in urban areas},
  author={Wang, Zhongqi and Han, Qi and De Vries, Bauke},
  journal={Applied Spatial Analysis and Policy},
  volume={15},
  number={4},
  pages={1193--1217},
  year={2022},
  publisher={Springer}
}

@article{qi2024understanding,
  title={Understanding the relationship between urban public space and social cohesion: A systematic review},
  author={Qi, Jie and Mazumdar, Suvodeep and Vasconcelos, Ana C},
  journal={International journal of community well-being},
  volume={7},
  number={2},
  pages={155--212},
  year={2024},
  publisher={Springer Science and Business Media LLC}
}

@article{he2026digital,
  title={Do digital platforms influence gentrification? An analysis of Nanjing’s central urban area},
  author={He, Qianhui and Sun, Shijie},
  journal={Environment and Planning B: Urban Analytics and City Science},
  pages={23998083261434364},
  year={2026},
  publisher={SAGE Publications Sage UK: London, England}
}

@article{gong2024google,
  title={Google effects on memory: a meta-analytical review of the media effects of intensive Internet search behavior},
  author={Gong, Chen and Yang, Yang},
  journal={Frontiers in public health},
  volume={12},
  pages={1332030},
  year={2024},
  publisher={Frontiers Media SA}
}

@article{wei2024deciphering,
  title={Deciphering the effect of user-generated content on park visitation: A comparative study of nine Chinese cities in the Pearl River Delta},
  author={Wei, Di and Wang, Yuan and Jiang, Yuxiao and Guan, Xueqing and Lu, Yi},
  journal={Land Use Policy},
  volume={144},
  pages={107259},
  year={2024},
  publisher={Elsevier}
}

@article{dong2023spatiotemporal,
  title={Spatiotemporal behavior pattern differentiation and preference identification of tourists from the perspective of ecotourism destination based on the tourism digital footprint data},
  author={Dong, Wei and Kang, Qi and Wang, Guangkui and Zhang, Bin and Liu, Ping},
  journal={Plos one},
  volume={18},
  number={4},
  pages={e0285192},
  year={2023},
  publisher={Public Library of Science San Francisco, CA USA}
}

@article{caprotti2022beyond,
  title={Beyond the smart city: A typology of platform urbanism},
  author={Caprotti, Federico and Chang, I-Chun Catherine and Joss, Simon},
  journal={Urban transformations},
  volume={4},
  number={1},
  pages={4},
  year={2022},
  publisher={Springer}
}

@article{du2022want,
  title={‘I want to record and share my wonderful journey’: Chinese Millennials’ production and sharing of short-form travel videos on TikTok or Douyin},
  author={Du, Xin and Liechty, Toni and Santos, Carla A and Park, Jeongeun},
  journal={Current Issues in Tourism},
  volume={25},
  number={21},
  pages={3412--3424},
  year={2022},
  publisher={Taylor \& Francis}
}

@article{koltai2025classifying,
  title={Classifying social position with social media behavioral data},
  author={Koltai, J{\'u}lia and Rakovics, Zs{\'o}fia and Kmetty, Zolt{\'a}n and Sz{\'a}mel, Kata and Ungvari, Borbala and Varadi, Bendeguz and Huszar, Akos},
  journal={EPJ Data Science},
  volume={14},
  number={1},
  pages={60},
  year={2025},
  publisher={Springer}
}

@article{wang2024multilingual,
  title={Multilingual e5 text embeddings: A technical report},
  author={Wang, Liang and Yang, Nan and Huang, Xiaolong and Yang, Linjun and Majumder, Rangan and Wei, Furu},
  journal={arXiv preprint arXiv:2402.05672},
  year={2024}
}

@article{healy2024uniform,
  title={Uniform manifold approximation and projection},
  author={Healy, John and McInnes, Leland},
  journal={Nature Reviews Methods Primers},
  volume={4},
  number={1},
  pages={82},
  year={2024},
  publisher={Nature Publishing Group UK London}
}

@article{mcinnes2017hdbscan,
  title={hdbscan: Hierarchical density based clustering.},
  author={McInnes, Leland and Healy, John and Astels, Steve and others},
  journal={J. Open Source Softw.},
  volume={2},
  number={11},
  pages={205},
  year={2017}
}

@misc{openstreetmapOpenStreetMap,
	author = {OpenStreetMap},
	title = {{O}pen{S}treet{M}ap --- openstreetmap.org},
	howpublished = {\url{https://www.openstreetmap.org/\#map=8/47.184/19.509}},
	year = {2026},
	note = {[Accessed 26-05-2026]},
}

@article{spyratos2017using,
  title={Using Foursquare place data for estimating building block use},
  author={Spyratos, Spyridon and Stathakis, Demetris and Lutz, Michael and Tsinaraki, Chrisa},
  journal={Environment and Planning B: Urban Analytics and City Science},
  volume={44},
  number={4},
  pages={693--717},
  year={2017},
  publisher={SAGE Publications Sage UK: London, England}
}

@misc{thinkwithgoogle,
	author = {Google},
	title = {thinkwithgoogle.com},
	howpublished = {\url{https://www.thinkwithgoogle.com/\_qs/documents/37/mobile-redefined-consumer-decision-shopper-journey-b.pdf}},
	year = {2016},
	note = {[Accessed 26-05-2026]},
}

@article{molitor2023digitizing,
  title={Digitizing local search: An empirical analysis of mobile search behavior in offline shopping},
  author={Molitor, Dominik and Daurer, Stephan and Spann, Martin and Manchanda, Puneet},
  journal={Decision Support Systems},
  volume={174},
  pages={114018},
  year={2023},
  publisher={Elsevier}
}

\end{document}


\title{Supplementary Materials}
\author{}
\maketitle

\section{Betweeenness Centrality}
\label{Betweeenness Centrality}
To further characterize the structural importance of clusters within the Collective Feedback Network through enhanced visual representation, betweenness centrality is computed for each node \(v \in V\). Betweenness centrality quantifies the extent to which a node functions as an intermediary bridge connecting different parts of the network through shortest paths.

For a node \(v\), betweenness centrality is defined as

\begin{equation}
BC(v)
=
\sum_{s \neq v \neq t}
\frac{
\sigma_{st}(v)
}{
\sigma_{st}
},
\end{equation}

where \(\sigma_{st}\) denotes the total number of shortest paths between nodes \(s\) and \(t\), and \(\sigma_{st}(v)\) represents the number of those paths passing through node \(v\).

To allow comparison across networks of different sizes, the measure is normalized as

\begin{equation}
BC^{\mathrm{norm}}(v)
=
\frac{
2 \times BC(v)
}{
(|V|-1)(|V|-2)
},
\end{equation}

where \(|V|\) is the total number of nodes in the Collective Feedback Network.

Within the Feedback Network framework, nodes with high betweenness centrality represent important bridging clusters connecting heterogeneous cognitive and spatial domains. A high centrality search cluster indicates that the semantic category plays a key intermediary role in linking multiple visitation activities, whereas a high centrality visitation cluster implies that the corresponding urban function acts as a major connector across diverse patterns of online search behavior. Accordingly, betweenness centrality captures the extent to which specific semantic or spatial clusters facilitate behavioral linkages and mediate the flow of cognitive-spatial interactions throughout the network.

\clearpage

\section{Hierarchical Dendogram}
\label{Hierarchical Dendogram}
To examine the similarity structure among clusters based on their concentration entropy characteristics, hierarchical agglomerative clustering was performed and visualized using a dendrogram. 

The similarity between two clusters \(c_p\) and \(c_q\) was quantified using the Euclidean distance metric. For the univariate case based solely on concentration entropy, the distance is defined as

\begin{equation}
d_{pq}
=
\left\lVert C_i^p - C_i^q \right\rVert_2
=
\sqrt{(C_i^p - C_i^q)^2}.
\end{equation}

An average linkage criterion was employed to determine the distance between two intermediate cluster groups. Let \(\mathcal{A}\) and \(\mathcal{B}\) denote two sets of clusters. The inter-group distance is computed as the arithmetic mean of all pairwise distances between elements belonging to the two groups:

\begin{equation}
D(\mathcal{A},\mathcal{B})
=
\frac{1}{|\mathcal{A}|\,|\mathcal{B}|}
\sum_{i \in \mathcal{A}}
\sum_{k \in \mathcal{B}}
d_{pq},
\end{equation}

where \(|\mathcal{A}|\) and \(|\mathcal{B}|\) represent the number of clusters contained in each group.

The agglomerative clustering procedure begins by treating each cluster as an independent unit. At each iteration, the pair of groups with the smallest average linkage distance is merged. This process continues iteratively until all clusters are combined into a single hierarchical structure.

The resulting dendrogram provides a graphical representation of the hierarchical similarity among clusters in heterogeneous cognitive and spatial domains according to their concentration entropy profiles. Clusters connected at lower dendrogram heights are characterized by search activities, whereas clusters merged at greater heights are associated with visit activities. The dendrogram therefore enables the identification of higher-order behavioral groupings and reveals latent organizational structures within the urban mobility system, suggesting a robust pattern of cognitive–spatial dualism.

\clearpage

\section{Distribution of Data}
\label{Distribution of Data}

\begin{figure}[!ht]
  \centering
  \includegraphics[width=1\textwidth]{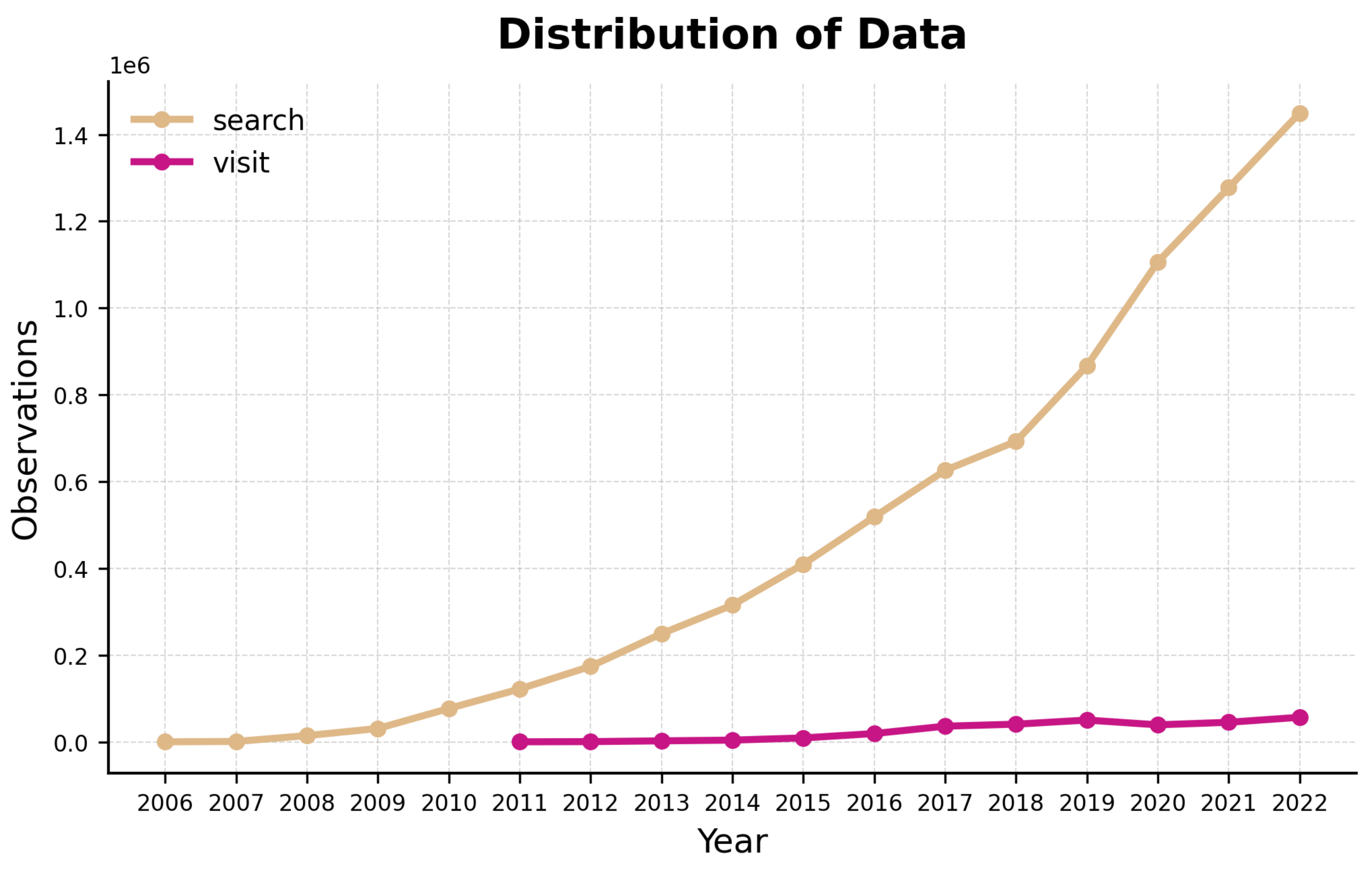}
  \caption{Distribution of the final dataset covering all data points obtained through data donation.}  
  \label{SM_B}
\end{figure}

\clearpage

\section{Spatial Distribution of Locations}
\label{Spatial Distribution of Locations}

\begin{figure}[!ht]
  \centering
  \includegraphics[width=1\textwidth]{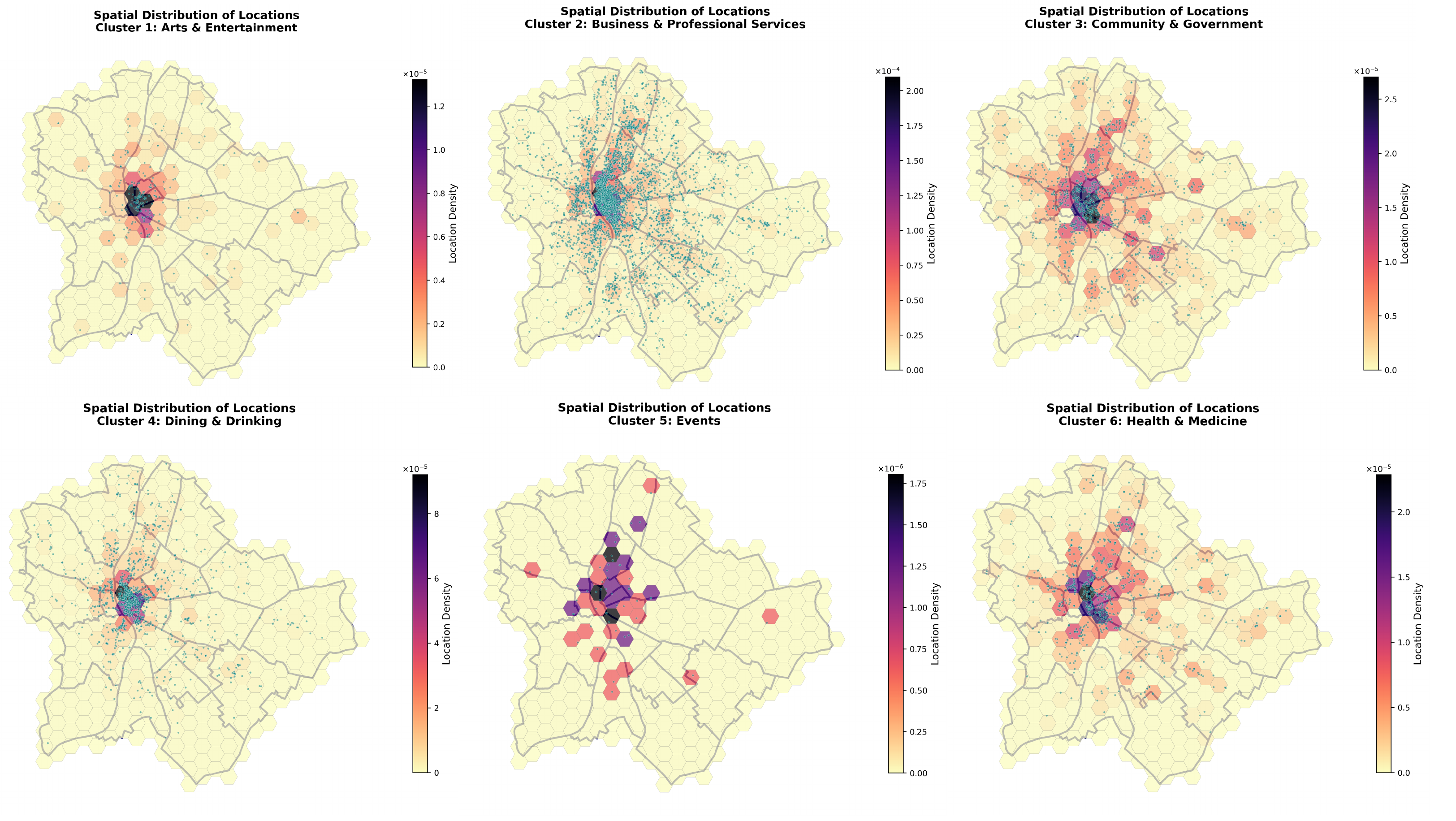}
  \caption{Spatial distribution of locations projected onto a uniform hexagonal grid with an 800-meter radius (Cluster 1-6).}  
  \label{SM_C1}
\end{figure}

\begin{figure}[!ht]
  \centering
  \includegraphics[width=0.7\textwidth]{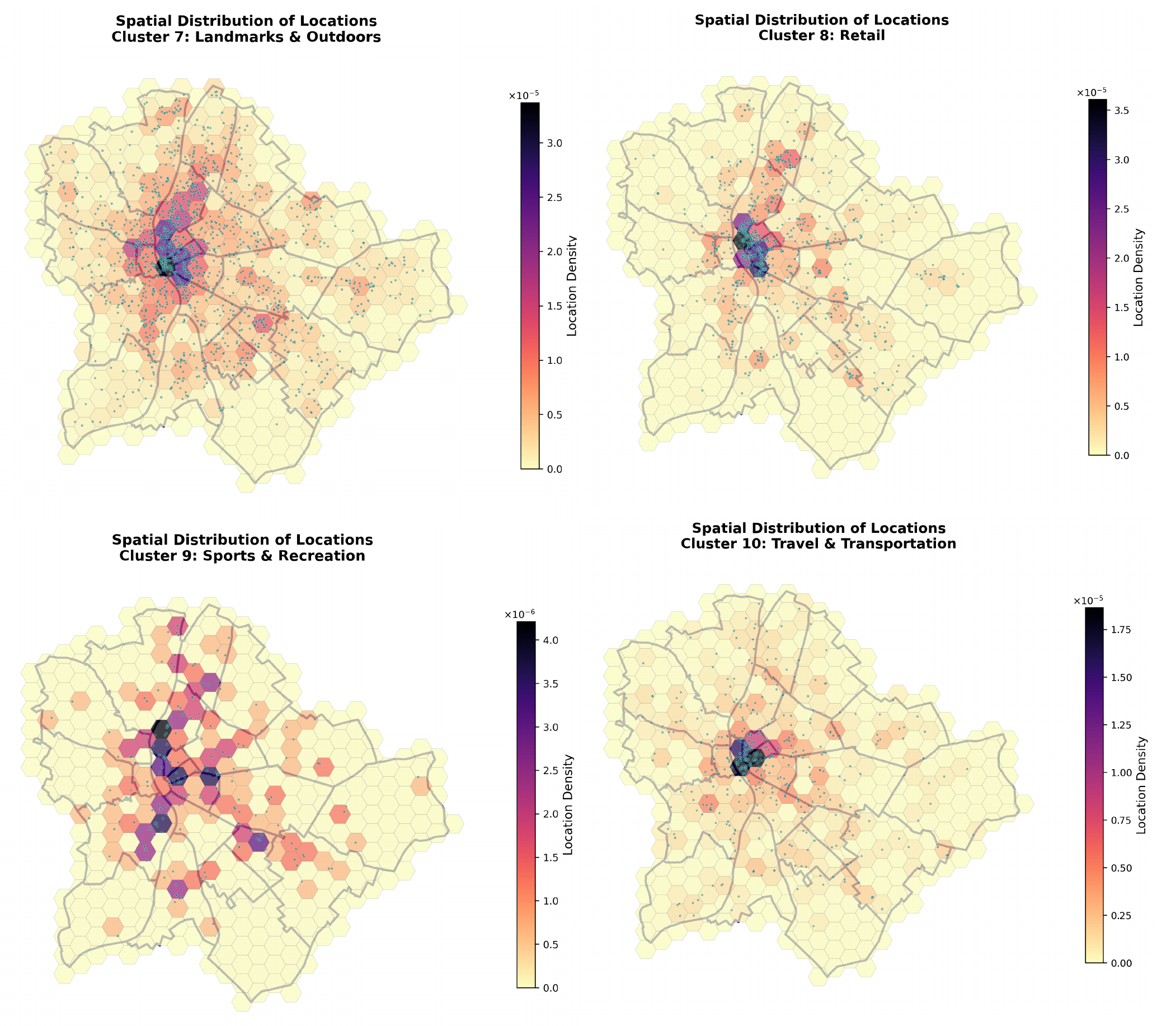}
  \caption{Spatial distribution of locations projected onto a uniform hexagonal grid with an 800-meter radius (Cluster 7-10).}  
  \label{SM_C2}
\end{figure}